\documentclass[12pt,journal]{IEEEtran}
\def\BibTeX{{\rm B\kern-.05em{\sc i\kern-.025em b}\kern-.08em
    T\kern-.1667em\lower.7ex\hbox{E}\kern-.125emX}}
\usepackage{balance}

\author{
    Sikai Yang\
    \texttt{syang126@ucmerced.edu}
    \\
    Miaomiao Liu\
    \texttt{mliu71@ucmerced.edu}
    \\
    Wan Du\
    \texttt{wdu3@ucmerced.edu}
    \\
}

\usepackage{eucal}
\usepackage{bm}
\usepackage{wrapfig}
\usepackage{makecell}
\usepackage{algorithm}
\usepackage{lineno}
\usepackage{algpseudocode}
\usepackage{amsmath}
\usepackage{subfigure}
\usepackage{booktabs}
\usepackage{multirow}
\usepackage{enumitem}
\usepackage{flowchart}
\usepackage{float}
\usepackage{times}
\usepackage{url}
\usepackage{amsfonts}

\usepackage{amssymb}
\usepackage{xspace}
\usepackage{graphicx}
\usepackage{tabu}
\usepackage{diagbox}
\usepackage{gensymb}
\usepackage{xcolor}
\usepackage{newtxtext,newtxmath}

\newcommand{\highlight}{\textcolor{black}}

\newcommand{\aliasSystem}{\textit{MDR}\xspace}
\begin{document}




\title{\LARGE Magnetic Distortion Resistant Orientation Estimation}

\maketitle
\pagestyle{plain}


\subsection*{Abstract}
Inertial Measurement Unit (IMU) sensors, including accelerometers, gyroscopes, and magnetometers, are used to estimate the orientation of mobile devices. 
However, indoor magnetic fields are often distorted, causing the magnetometer's readings to deviate from true north and resulting in inaccurate orientation estimates. 
Existing solutions either ignore magnetic distortion or avoid using the magnetometer when distortion is detected.
In this paper, we develop \aliasSystem, a Magnetic Distortion Resistant orientation estimation system that fundamentally models and corrects magnetic distortion. 
\aliasSystem builds a database to record magnetic directions at different locations and uses it to correct orientation estimates affected by magnetic distortion.
To avoid the overhead of database preparation, \aliasSystem adopts practical designs to automatically update the database in parallel with orientation estimation.
Experiments on 27+ hours of arm motion data show that \aliasSystem outperforms the state-of-the-art method by 35.34$\%$. 

\section{Introduction}
\begin{figure}[h]
    \hfill
    \begin{minipage}[t]{0.4\linewidth}
        \centering
        \includegraphics[width=\linewidth]{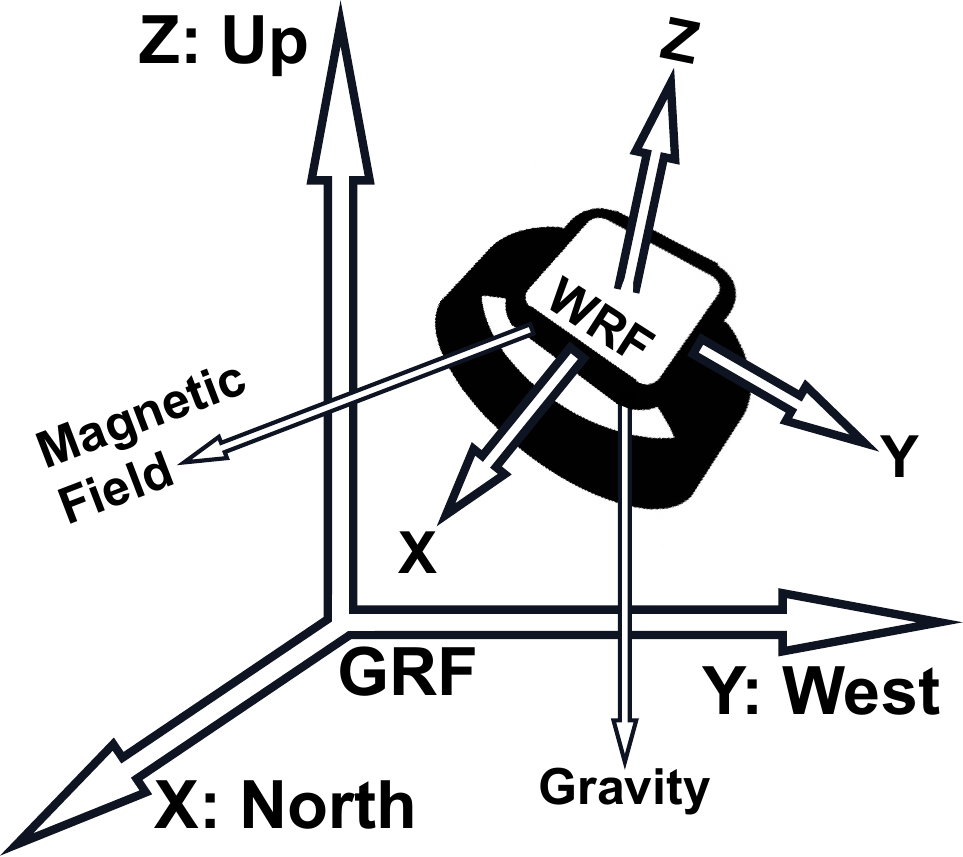}
        \vspace{-5mm}
        \caption{WRF and GRF.}
        \label{W&G}
    \end{minipage}
    \hfill
    \vspace{-3mm}
\end{figure}

Arm tracking is essential in many applications, such as gesture recognition~\cite{shen2016smartwatch, ren2021winect, chen2021rf, zhang2019towards, ma2021location}, Virtual Reality (VR) \cite{VR}, and smart healthcare~\cite{zhao2019ultigesture,zhao2018mobigesture, zhao2017gesture, kong2022m3track}.
With the pervasive acceptance of smartwatches, these devices provide a convenient and ubiquitous way for arm tracking. 
Inertial Measurement Unit (IMU)~\cite{jung2021lax, dai2023interpersonal} sensors (i.e., gyroscope, accelerometer, and magnetometer) in smartwatches measure and track the orientation and location of the user's wrist. 
We focus on orientation estimation, which is a prerequisite for location estimation in arm tracking \cite{muse,liu2019real, gao2022mom}, as well as smart applications like in-air writing and star sky rendering.
Orientation refers to the relationship between the Watch's Reference Frame (WRF) and the Global Reference Frame (GRF). 
The WRF aligns with the measurement axes of the IMU sensors, while the GRF is established with north as the X-axis, west as the Y-axis, and up as the Z-axis, as shown in Figure \ref{W&G}.

Existing orientation estimation methods assume that the magnetic field directions are constant. 
However, magnetic distortion causes these directions to vary significantly. 
For example, we find an average magnetic field direction deviation of 31° in a corridor. 
This leads to substantial errors in magnetic calibration and, consequently, orientation estimation.

\highlight{
The state-of-the-art work, MUSE \cite{muse}, uses a magnetometer for constant calibration. 
It employs a complementary filter to calibrate orientation tracking results using magnetometer, and also accelerometer when the device is static.
However, MUSE fails to handle the magnetic distortion commonly found in buildings, where metal structures cause the magnetic field to point in various directions~\cite{kim2016novel, zhang2017deeppositioning, shu2015magicol, wang2021mvp, xie2014maloc, ho2020using, shao2016location, DEVRIES2009535, chung2011indoor, ahrs}.
}
According to experiments \cite{muse}, MUSE's error could rise up to 100° in the presence of magnetic distortion.
Current distortion handling methods \cite{fan2017adaptive, ahrs, madgwick} either reduce or avoid using the magnetometer when detecting magnetic distortion.
\highlight{They use the intensity and direction of the magnetic field to indicate distortion intensity
Based on this assessment, they adjust the weight of magnetometer usage.
}
While they partially mitigate the impact of magnetic distortion, they also suffer from the absence of magnetometer calibration, thus fail when distortion is high.




We develop \aliasSystem, a Magnetic Distortion-Resistant orientation estimation system that models and resists magnetic distortion in orientation estimation. 
\aliasSystem builds a database for modeling magnetic distortion to protect orientation estimation from magnetic distortion. 
The database divides the 3D space into voxels of 0.001$m^3$ to accurately model the magnetic distortion.
For orientation estimation, \aliasSystem utilizes the complementary filter, similar to the prior work MUSE. 
However, rather than assuming a constant magnetic field direction, we query our database to determine the actual magnetic direction in the GRF.
We use this magnetic direction provided by the database to perform magnetic calibration.
This involves rotating the estimated WRF so that the magnetometer reading (measured in WRF) aligns with the actual magnetic field direction (in GRF), thereby calibrating the orientation result.
Yet, to make \aliasSystem practical, we address three major challenges with corresponding designs:

\highlight{}
\textit{Automatic Database Construction:} 
Preparing the database can be challenging, requiring measurement of the magnetic field direction at each location. 
For instance, with a spatial resolution of 0.1m, filling a $1m^3$ space would necessitate 1000 measurements.
Lower resolution would fail to model the magnetic distortion.
Moreover, even after preparing a database, accurately utilizing it poses difficulties, as the device's exact location within the database's coordinate system is typically unknown. 
While IMU-based location estimation can track a device's relative movement, determining its absolute position within a room or another coordinate system, such as the database, remains challenging.
To mitigate these challenges and adapt autonomously to varying environments, \aliasSystem constructs and updates its database using estimated orientation and location data, in parallel with orientation estimation.
Using estimated orientation, \aliasSystem translates magnetometer readings measured in WRF into magnetic field directions in GRF. 
It then updates the corresponding voxel in the database based on the estimated location.

\textit{Adaptive Database Updating:}
Database updating relies on orientation estimation results, whose error may impact the updating accuracy of the database.
To address this, we introduce an adaptive updating scheme to enhance the quality of database construction.
It mainly uses the gyroscope as reference to assess the accuracy of estimated orientation.
Based on that, we implement weighted updating for the database to mitigate the impact of orientation results that may have high error.
Experiments indicate that the adaptive updating scheme decreases database error by 9\%.

\textit{Magnetic Distortion Detection:}
As the database is specifically designed to resist distortion, it does not contribute when distortion is absent. 
Therefore, it is practical to deactivate the database during mild distortion to conserve computational resources. 
For example, in a conference room where the distortion level is similar to outdoor, we observed that using the database does not enhance orientation estimation accuracy at all. 
To identify these distortion-free scenarios, we devise two criteria that, when used together, achieve 96\% accuracy.

We collect 27+ hours of arm motion data from 12 volunteers, at 10 places in two cities.
Experiments show that \aliasSystem improves orientation accuracy by 35.34$\%$ compared to the state-of-the-art method, MUSE.
We further conduct two real application case studies to demonstrate the direct improvement on smart applications enabled by \aliasSystem.

In summary, this paper makes the following contributions:


\begin{itemize}
    \item We develop \aliasSystem, a magnetic distortion-resistant orientation estimation system. 
    To the best of our knowledge, this is the first work that fundamentally models and resists magnetic distortion in orientation estimation.
    \item An automatic database construction method designed to make \aliasSystem auto-adapt to new scenarios.
    
    \item An adaptive updating scheme to mitigate the impact of orientation error and improve database accuracy.

    \item A distortion detection module that ensures the database is only activated when necessary.
    

    \item Extensive experiments on 27+ hours of data to evaluate \aliasSystem and baseline methods, and two case studies to demonstrate the direct improvement of \aliasSystem.

    
\end{itemize} 
\vspace{-3mm}

\section{Related Work}\label{related}
\textbf{Magnetic Fingerprint:}
Magnetic fingerprinting has been well-studied in localization \cite{davidson2016survey, kim2016novel, gozick2011magnetic, zhang2017deeppositioning, shu2015magicol, wang2021mvp, xie2014maloc, ho2020using, shao2016location, chung2011indoor, chung2011indoor, nagai2019indoor}.
While distortion is harmful in orientation estimation,  it provides valuable information in localization.
Due to magnetic distortion, the magnitude distribution of magnetic field varies in space.
Users can then localize themselves via matching their magnetometer readings with the pre-scanned magnetic fingerprints.
However, handling distortion in orientation estimation is different from fingerprinting-based localization.


Magnetic fingerprints are usually measured with a resolution of meters.
For example, MVP \cite{wang2021mvp} locates cars in tunnels using magnetic fingerprint and has a distance interval of 3.89 m.
However, resisting distortion in orientation estimation requires finer spatial resolution.
According to our experiments in Section \ref{secDatabaseResolution}, the optimal resolution for our arm tracking task is 0.1m.
Fine resolution increases the workload of database preparation, making it almost impossible to pre-scan a database manually.
\highlight{
To that end, we propose a novel automatic database updating mechanism (Section \ref{db_updating}). 
}

\highlight{
\textbf{IMU-based Orientation Estimation:} 
Compared to other computer vision or wireless sensing-based works, IMU-based orientation estimation has an advantage of independence, as it only requires the IMU sensors on the mobile platform.
}

Many sensor fusion algorithms are designed to process IMU sensor data for orientation estimation \cite{A3, muse}, such as Kalman Filter \cite{jurman2007calibration, yean2016algorithm, de2011uav, huyghe20093d}, Extended Kalman Filter \cite{marins2001extended, gebre2004design}, and Complementary Filter \cite{madgwick2011estimation, muse}. 
However, Kalman Filter-based methods assume that the error model is Gaussian. 
This assumption may easily break down in the context of human arm motion, as verified by MUSE~\cite{muse}.
Thus, like MUSE, \aliasSystem adopts a complementary filter.

Recent solutions have also explored data-driven methods for IMU orientation estimation \cite{miao,esfahani2019orinet, brossard2020denoising, sun2021idol}.
However, these methods require collecting a huge amount of data to support accurate estimation.
Additionally, data-driven methods may fail to generalize to new environments,
where the data distribution has not been seen in training data.

\textbf{Handling Magnetic Distortion in Orientation Estimation:}
There are few works that address magnetic distortion in IMU-based orientation estimation.
Earth magnetic field typically has a magnitude around $45\mu T \sim 60\mu T$, and forms a certain angle with the gravity direction, mainly depending on the latitude.
Leveraging this characteristic, three works \cite{fan2017adaptive, ahrs, madgwick} propose to use the magnitude as indicator of magnetic distortion. 
Additionally, two of them \cite{fan2017adaptive, ahrs} incorporate the dip angle of the magnetic field as an additional indicator.
They decrease the weight for magnetometer as distortion rises, thus avoiding the negative impact of magnetic distortion.
\highlight{
However, if distortion is intense, these methods assign relatively low weight to the magnetometer, leading to issues due to either the absence of magnetic calibration or the presence of magnetic distortion.
Instead, in this paper, we propose a novel system that models and corrects magnetic distortion in orientation estimation, even when distortion is high.
}
Other works use artificially generated magnetic fields to track devices but require additional hardware \cite{miposer}.
\section{Background} 
\label{secBackground}

\aliasSystem uses a smartwatch as platform and IMU sensors as input.
It uses a complementary filter for orientation estimation. 
To model the magnetic distortion, it needs to infer location.
To that end, \aliasSystem use a particle filter, similar to MUSE \cite{muse}.
In this section, we briefly introduce the related background.

\textbf{IMU Sensors for Orientation Estimation:}
Orientation refers to the relationship between Watch's Reference Frame (WRF) and Global Reference Frame (GRF), as shown in Figure \ref{W&G}.
The WRF axes are either parallel or orthogonal to the watch's edges, and IMU sensor readings are all measured in WRF.
If we know the angular relationships between the axes of WRF and GRF, we can describe the watch’s orientation in GRF. For example, a mobile device can measure north using its magnetometer; and it can also measure the gravity using accelerometer when it is static or moving at a constant speed.
Based on these two GRF directions measured in WRF, we can determine the relationship between WRF and GRF, i.e., the watch's orientation in GRF.


However, when the arm is in motion, accelerometer does not accurately measure the gravity direction because it senses both the gravitational acceleration and the linear acceleration of arm movement.
Therefore, we add gyroscope to track the mobile device's orientation. 
Gyroscope measures angular velocity.
Based on an initial orientation, we perform discrete integration on the angular velocity to track orientation.
However, any errors in gyroscope readings will accumulate during integration, leading to drift in orientation estimation.
Without frequent calibrations, the gyroscope based orientation tracking will drift away from the true orientation. 
Therefore magnetometer comes in and provides an extra direction for calibration.
However, the magnetic field can be distorted by metal objects, causing the magnetometer to point to incorrect directions, which will be discussed further in this work.


\textbf{Orientation Estimation via Complementary Filter:}
Complementary filter first initializes an orientation using accelerometer and magnetometer when the device is static.
Based on the initial orientation, \aliasSystem iteratively integrates the gyroscope reading to track orientation, which generates a rotation matrix $R_{g}$ at each time step.
However, errors in gyroscope integration causes a drift in the orientation result, which can be calibrated via accelerometer and magnetometer.
To use the accelerometer for calibration, we first define 'down' direction in GRF as the gravity anchor.
With current orientation estimation result, we can transform the accelerometer measurement from WRF into GRF.
Now, we have two vectors in GRF, i.e., the transformed accelerometer direction and the gravity anchor.
By comparing them, we can have a rotation matrix $R_{a}$ that calibrates the transformed accelerometer direction to the gravity anchor.
Meanwhile, magnetometer measures the Earth magnetic field, which is supposed to point north if there is no magnetic distortion.
Similar to accelerometer calibration, we use the north direction in GRF as the magnetic anchor, to calculate the rotation matrix $R_{m}$ that calibrates the transformed magnetometer direction to the magnetic anchor.
Based on the above three rotations, the complementary filter fuses them to update orientation:
\begin{equation}
    \Theta(t+\Delta t)=\Theta(t) \cdot R_{g} \cdot R_{a}(k_{a}) \cdot  R_{m}( k_{m}),
    \label{fuse}
\end{equation}
Here, $R_{g}$, $R_{a}$, and $R_{m}$ represent the rotation matrix derived from gyroscope-based orientation tracking, accelerometer calibration and magnetometer calibration respectively. 
$k_{a}$ and $k_{m}$ are two weight parameters, controlling the percentage of the rotation we perform by accelerometer calibration and magnetometer calibration.
Since complementary filter is an iterative process (i.e., 50 rounds per second in our system setting), accelerometer calibration and magnetometer calibration can be performed gradually with small weights in Equation \ref{fuse}.
Based on extensive experiments, we set both $k_{a}$ and $k_{m}$ as 0.1 for the best orientation estimation performance.



\highlight{
\textbf{Magnetic Anchor:}
}
\highlight{
Magnetic anchor represents the target direction of magnetic calibration.
As introduced above, magnetic calibration is an important procedure in IMU sensor fusion.
It requires two pieces of information: magnetometer measurement (in WRF), and magnetic anchor (in GRF).
While existing methods usually regard the magnetic anchor as a constant direction in GRF, the true magnetic anchor is the actual direction of the magnetic field, which may be different at different locations due to the existence of distortion.
}



\textbf{Location Estimation via Particle Filter:} \label{particle}
Database operations require location information.
\highlight{While many studies have been done on location estimation of arm tracking \cite{cutti2008ambulatory,el2012shoulder,riaz2015motion, tautges2011motion, shen2016smartwatch, liu2019real, muse}, \aliasSystem follows the state-of-the-art arm tracking solution, MUSE, and adopts the particle filter to estimate the smartwatch's location.}
Here, we briefly introduce it.
Due to the constraints of human skeleton, the space of possible smartwatch locations is limited for specific wrist orientations.
The particle filter randomly samples a set of particles onto the possible locations.
Based on the orientation result at each time step, it samples new locations for the particles, which forms a trajectory.
The particle filter uses accelerometer to
evaluate how well each particle tracks the motion.
It then discards bad particles and regenerates new particles around the survivors. 
As this process repeats, the particles are continuously refined to track the device's location.

\section{MDR Design} \label{design}
In this section, we introduce the design of \aliasSystem, including three key components: database query, automatic updating, and distortion detection.

\begin{figure}[h]
    \vspace{-1mm}
    \hfill
    \begin{minipage}[t]{0.45\linewidth}
        \centering
        \includegraphics[width=\linewidth]{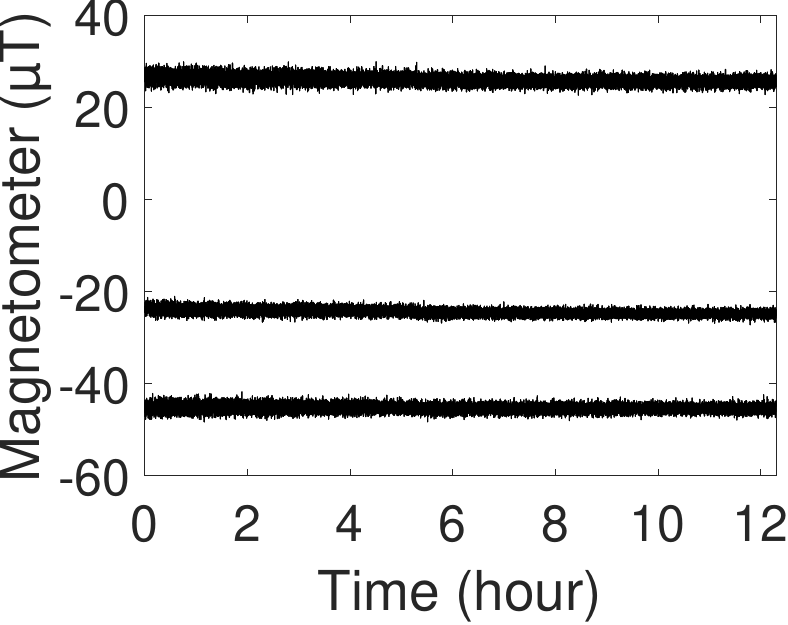}
         \vspace{-5mm}
        \caption{Temporal stability of the magnetic field.}
        \label{temporal}
    \end{minipage}
    \hfill
    \begin{minipage}[t]{0.45\linewidth}
        \centering
        \includegraphics[width=\linewidth]{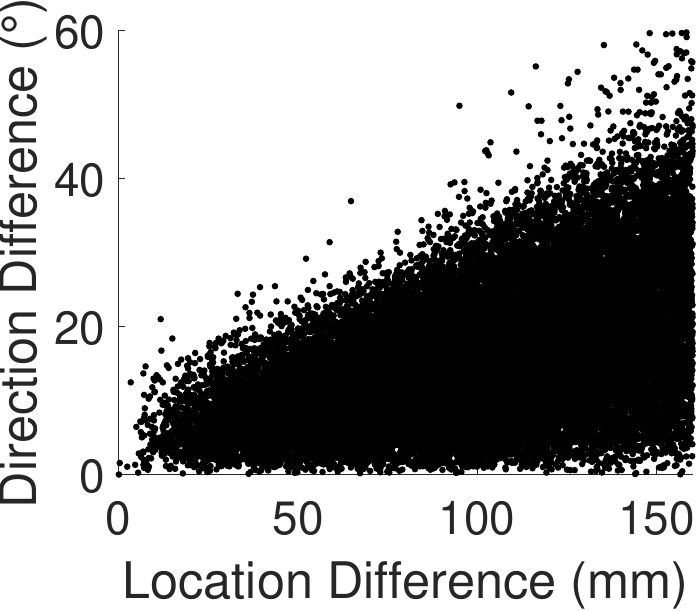}
         \vspace{-5mm}
        \caption{Spatial correlation of the magnetic field.}
        \label{spatial}
    \end{minipage}
    \hfill
    \vspace{-3mm}
\end{figure}

\subsection{Analysis on Magnetic Distortion} 
\label{DBdesign}
We first study the nature of magnetic distortion, mainly from temporal and spatial aspects.
These observations motivates us to build a database to model distortion.

\textbf{Temporal Stability:}  \label{stability}
Geological magnetic fields are stable, and artificial metal structures remain stationary, resulting in a consistent distortion pattern of the magnetic field.
Figure \ref{temporal} shows the 3-axis readings of a static magnetometer in an indoor environment over 12 hours.
Several other works on magnetic fingerprinting also verify this chronological stability \cite{chung2011indoor,xie2014maloc,shao2016location,wang2021mvp}. 
In conclusion, in most cases, we can regard the magnetic field as a constant field.
This temporal stability validates the feasibility of modeling distortion and using it to assist algorithms when devices revisit past locations.

\textbf{Spatial Correlation:}
\label{spatial corr}
We then explore the spatial correlation of magnetic field.
If the distance is limited, the difference of magnetic field directions is also limited, as shown in Figure \ref{spatial}.
This spatial correlation suggests that we can model the distortion with certain spatial resolution.

\subsection{\aliasSystem overview} \label{DB}
\highlight{
\textbf{Magnetic Distortion Database}:
}
To model magnetic distortion, \aliasSystem builds a magnetic distortion database that divides the 3D space into voxel datapoints with a resolution of $l_{DB}$.
Based on experiments, we set $l_{DB}=0.1m$.
Each datapoint stores the magnetic field direction measured in GRF (i.e., magnetic anchor) at that location.
The database shares the coordinate system established by the complementary filter during initialization.
Database operations mainly include two components: query, updating.

\highlight{
\textbf{Database Query} (\textit{Section \ref{deadlock}})\textbf{:} 
}
At each time step, \aliasSystem queries the database using estimated location to acquire an magnetic anchor.
This anchor is then fed into the magnetic calibration process for orientation estimation.
As the location is inferred via a particle filter using the orientation result, we encountered and resolved a deadlock issue to ensure the practicality of the query process.

\highlight{
\textbf{Automatic Updating} (\textit{Section \ref{selfDB}})\textbf{:}
}
Database preparation is a challenging task.
Since the database stores the magnetic field direction at various locations, supportive systems for orientation and location estimation may be required for database preparation (e.g., cameras, wireless sensing).
To eliminate this overhead and auto-adapt to new places, \aliasSystem utilizes estimated orientation and location
to automatically build and update the database in parallel.

\textbf{Adaptive Updating} (\textit{Section \ref{adaptive_updating}}):
Updating the database relies on orientation results, which may introduce errors affecting database accuracy.
To that end, we design an adaptive updating scheme to reduce the updating weight when the orientation result may have high error.
It uses the gyroscope to assess the credibility of estimated orientation.
Based on the assessment, we further develop an IAI (Inertial Angular Index) to better capture the relationship between gyroscope and potential orientation error.

\highlight{
\textbf{Distortion Detection} (\textit{Section \ref{determine use or not}})\textbf{:}
}
The database is specifically designed to handle distortion and does not contribute when distortion is absent. 
To optimize computation, \aliasSystem employs two criteria to detect distortion-free locations. 
If distortion is mild or absent according to these criteria, the system automatically turns off the database.

\begin{figure}[h]
    \vspace{-1mm}
    \hfill
    \begin{minipage}[t]{0.78\linewidth}
        \centering
        \includegraphics[width=\linewidth]{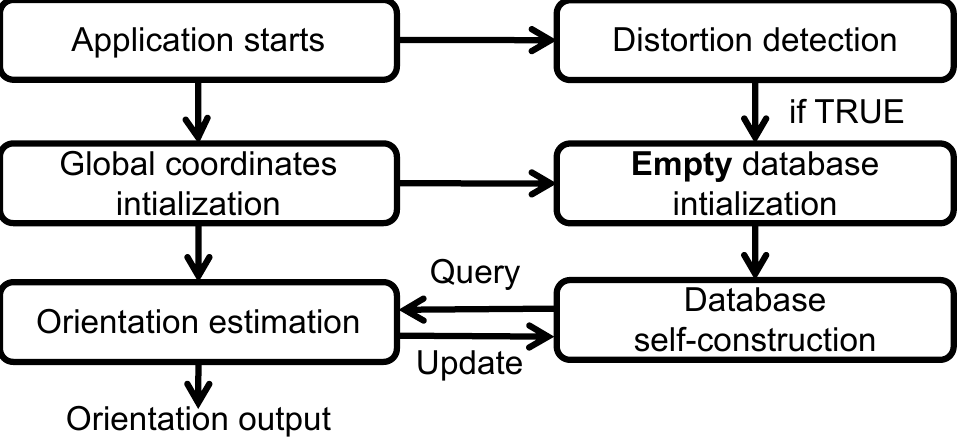}
         \vspace{-5mm}
        \caption{MDR Application Workflow.}
        \label{app_flow}
    \end{minipage}
    \hfill
    \vspace{-3mm}
\end{figure}

\highlight{
\textbf{Application Workflow:} \label{sec_app_flow}
}
We demonstrate the practical application of \aliasSystem in Figure~\ref{app_flow}. 
The workflow begins when a user launches an application requiring orientation estimation, such as in-air writing or virtual sports. 
The user is instructed to remain still for the first 10 seconds to initialize the Gravity Reference Frame (GRF) for accurate orientation tracking. 
After that, if magnetic distortion is detected, an empty database is initialized.
After initialization, the orientation estimation algorithm begins working, and the database performs 'Query' and 'Update' operations based on sensor measurements to construct itself.
When the user stops the application, \aliasSystem does not need to save the database but builds a new one next time it is used.
In any new scenario, \aliasSystem relies on automatic updating to explore and adapt without any pre-modelling.

\subsection{Database Query} \label{deadlock}

\aliasSystem queries the database to acquire a magnetic anchor and feeds it to the complementary filter for correct magnetic calibrations. 
If no anchor exists at the queried datapoint, \aliasSystem disables magnetic calibration in the complementary filter for that time step. 
Upon revisiting the same datapoint, an anchor will be available for query due to the updating mechanism.

\begin{figure}[ht]
    \centering
    \includegraphics[width=0.99\linewidth]{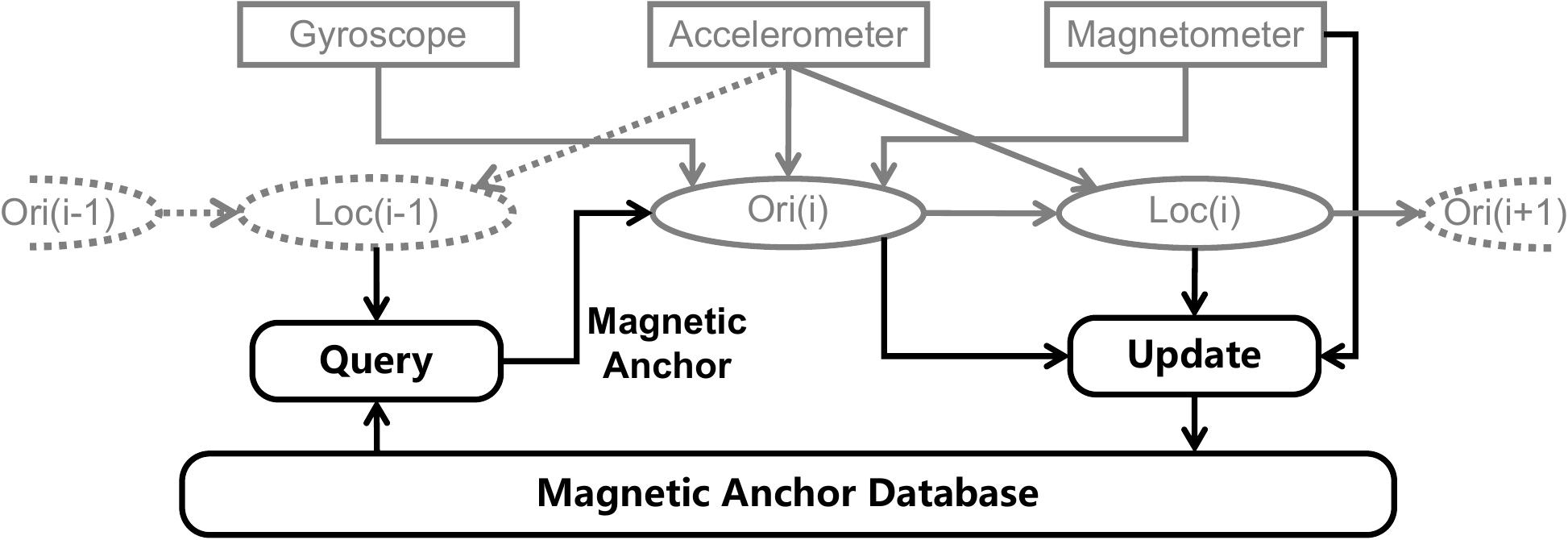}
    \vspace{-3mm}
    \caption{MDR System Workflow.}
    \label{flow}
    \vspace{-2mm}
\end{figure}

\textbf{System Deadlock:}
However, a deadlock issue arises: the particle filter requires orientation to update the location, while the complementary filter needs an anchor from the database to estimate orientation. 
Additionally, the database requires a location result from the particle filter for querying.
As a result, \aliasSystem encounters a deadlock.

An intuitive solution to the deadlock problem is to temporarily use the complementary filter for orientation estimation without relying on the database. 
This temporary orientation estimate can then be used by the particle filter to infer a temporary location. 
Subsequently, the temporary location is used to query the database and retrieve a magnetic anchor, which is fed back into the complementary filter for the final orientation estimation. 
This approach involves running the complementary filter twice within one time step, and the temporary orientation generated by the first complementary filter may not be accurate.

Instead, we propose another solution that utilizes location results from previous time steps to achieve better accuracy and efficiency. 
Figure \ref{flow} illustrates the system workflow based on this new approach. 
This method leverages the continuity of human arm motion, where the limited speed of human motion and the short interval of smartwatch sensor data sampling (20ms) constrain the possible distance between two consecutive time steps.
Based on data collected from 12 users, we examined the distance between consecutive time steps. 
The results show that the average distance between two consecutive time steps is 7.6mm, and more than 98.5\% of these distances are below 20mm.
Covering a distance of 20mm within 20ms indicates a speed of 1m/s, which is unusually fast for daily arm motions. 
From the spatial distribution of magnetic distortion shown in Figure \ref{spatial}, it is evident that a 20mm movement does not significantly alter the direction of the magnetic field. 
Therefore, we can effectively use the location result from the previous time step for database queries
Experiments demonstrate that this continuity-based solution has 8.36\% less error compared to the initial approach and is more efficient. 
As a result, \aliasSystem adopts this solution to resolve the deadlock problem.

\subsection{Automatic Database Updating} \label{selfDB}

Database preparation can be challenging, requiring numerous artificial measurements to fill the 3D space for modeling the magnetic field. Even after preparation, using the database poses difficulties because IMU-based localization can only track the trajectory, not the absolute location of the device within the database's coordinates.

To alleviate the overhead of database preparation, we propose building the database concurrently with the orientation estimation process. \aliasSystem automatically constructs and updates the database based on orientation and location results. This approach ensures that device localization and database construction share the same coordinate system.

\label{db_updating}
\highlight{
\aliasSystem uses three information pieces at each time step to update the database: magnetometer reading $\overrightarrow{M}$, estimated orientation $\hat{\Theta}$, and the estimated location from particle filter $\overrightarrow{x}$.
With the magnetometer reading $\overrightarrow{M}$ measured in WRF and the estimated orientation $\hat{\Theta}$ of the watch, \aliasSystem first calculates the magnetic anchor $\overrightarrow{N}$ in GRF: 
}
\highlight{
\begin{equation}
    \overrightarrow{N} = \overrightarrow{M}\cdot\hat{\Theta}
    \label{Fanchor}
\end{equation}
}
\aliasSystem then stores the anchor $\overrightarrow{N}$ at location $\overrightarrow{x}$ in the database.
Consequently, an initially empty database gradually accumulates magnetic anchors at each datapoint.
Since a smartwatch may revisit a datapoint multiple times, the final result queried from the database uses the average direction of all anchors at that datapoint.
\highlight{
The updating formula is shown below, where $\overrightarrow{DP}_{old}$ is the old datapoint in the voxel before updating, and $\overrightarrow{DP}_{new}$ is the updated datapoint:
}
\highlight{
\begin{equation}\label{Fupdate}
    \overrightarrow{DP}_{new}=\overrightarrow{DP}_{old}+\overrightarrow{N}
\end{equation}
}
Averaging multiple anchors will reduce their overall error, gradually enhancing the database accuracy.

\subsection{Adaptive Database Updating}
\label{adaptive_updating}
\aliasSystem can now build and update its own database, but the magnetic anchors it stores may not always be accurate.
As shown in Equation \ref{Fanchor}, a magnetic anchor is calculated using the magnetometer reading $\overrightarrow{M}$ and the estimated orientation $\hat{\Theta}$.
Errors in $\hat{\Theta}$ will affect the accuracy of the magnetic anchors stored in the database.
To mitigate the impact of orientation errors on database accuracy, we propose an adaptive updating scheme.
This scheme grants lower weights to the anchors if the orientation results are deemed untrustworthy. 
The key challenge lies in quantifying the trustworthiness of an orientation result.


\begin{figure*}[ht]
    \hfill
    \begin{minipage}[t]{0.235\linewidth}
        \centering
        \includegraphics[width=\textwidth]{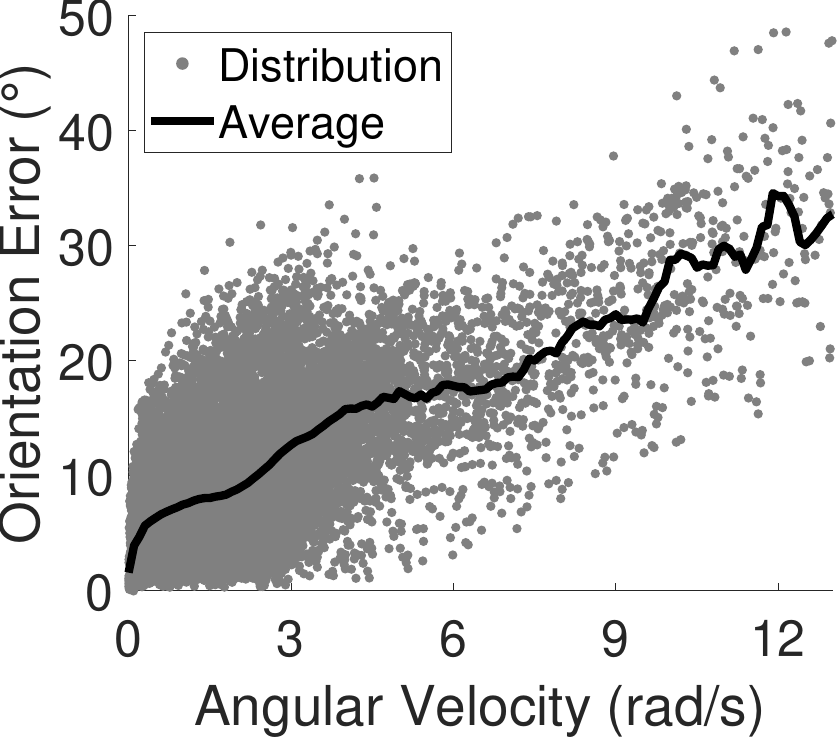}
         \vspace{-0.65cm}
        \caption{Relationship between angular velocity magnitude and orientation error.}
        \label{gyro_ori}
    \end{minipage}
    \hfill
    \begin{minipage}[t]{0.235\linewidth}
        \centering
        \includegraphics[width=\textwidth]{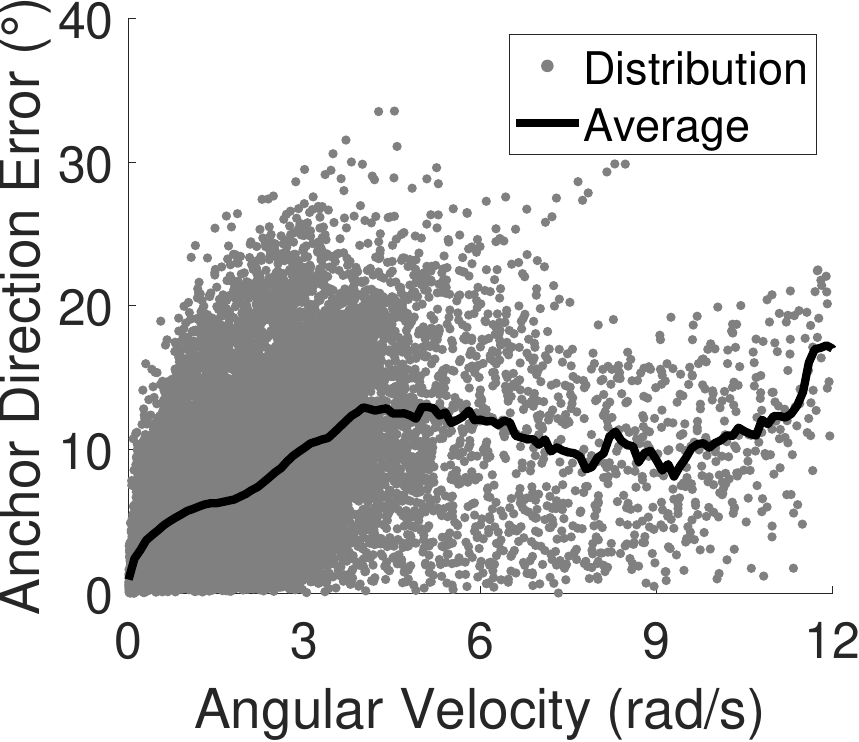}
         \vspace{-0.65cm}
        \caption{Relationship between angular velocity magnitude  and anchor error.}
        \label{gyro_anchor}
    \end{minipage}
    \hfill
    \begin{minipage}[t]{0.235\linewidth}
        \centering
        \includegraphics[width=\textwidth]{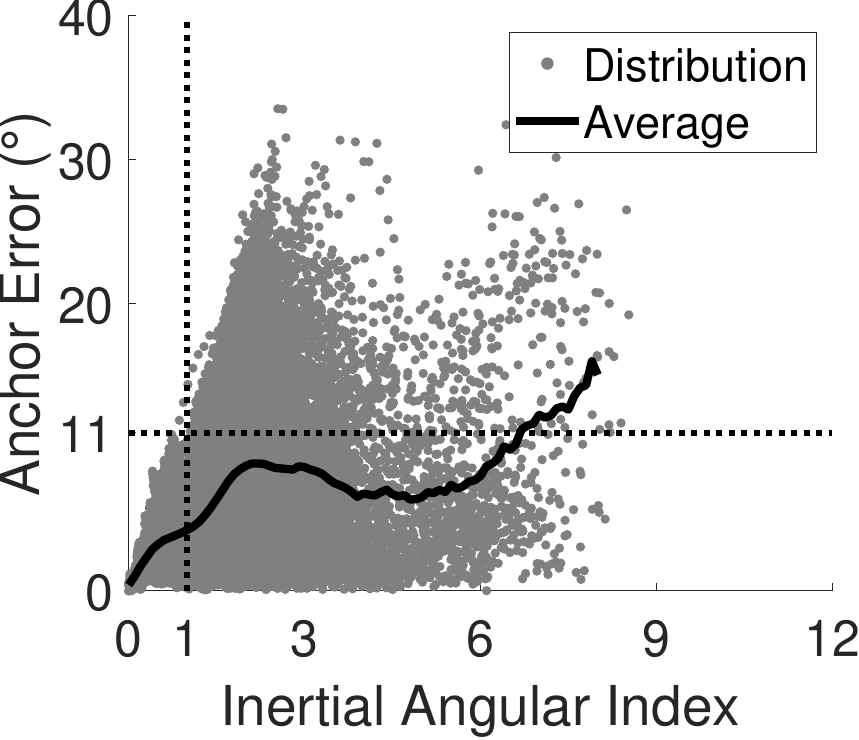}
         \vspace{-0.65cm}
        \caption{Relationship between Inertial Angular Index (IAI) and anchor error.}
        \label{IAI_anchor}
    \end{minipage}
    \hfill
    \vspace{-2mm}
\end{figure*}

We use the magnitude of angular velocity, specifically the gyroscope reading, as a reference to predict the accuracy of the orientation result. 
The gyroscope plays a significant role in orientation estimation. 
As shown in Equation \ref{fuse}, in IMU sensor fusion, gyroscope typically receives full weight, while magnetometer and accelerometer are granted relatively low weights.
When the magnitude of angular velocity is high, the error of the gyroscope also tends increase~\cite{A3}, leading to higher errors in the estimated orientation through gyroscope integration. 
Based on real experiments on the \aliasSystem prototype, which integrates all components up to this section, the distribution of orientation errors is depicted in Figure \ref{gyro_ori}.
Orientation errors contribute to errors in anchor directions, although they are not necessarily identical. 
Therefore, we also examine the relationship between anchor direction errors and the magnitude of angular velocity, as shown in Figure \ref{gyro_anchor}.

\textbf{Inertial Angular Index:}
However, we do not directly use angular velocity as the index for adaptive updating.
Since angular velocity is integrated over time to track orientation, a high magnitude does not instantly guarantee high orientation error.
It is more reasonable to monitor the cumulative magnitude of angular velocity over a short period.
Therefore, we propose an Inertial Angular Index, denoted as $IAI$: 
\begin{equation}
    IAI(t)=IAI(t-\Delta t)*k_{IAI}+\omega(t)*(1-k_{IAI}) 
\end{equation}
Here, $\omega(t)$ represents the magnitude of angular velocity measured at time $t$, and $k_{IAI}$ is an inertial coefficient ranging from 0 to 1.
$IAI$ starts at 0 and accumulates the magnitude of angular velocity over time. 
This index reflects the overall magnitude of angular velocity within a short period while remaining computationally efficient.
Figure \ref{IAI_anchor} illustrates the relationship between $IAI$ and anchor error, showing a more linear trend compared to the relationship between gyroscope and anchor error in Figure \ref{gyro_anchor}. 
Consequently, we cap the anchor error within 11° when $IAI \leq 1$.
With $IAI$ established, we refine the updating scheme. 
Unlike Equation \ref{Fupdate}, we introduce a threshold $IAI_0$ for the $IAI$.
During updating, if $IAI$ exceeds this threshold, the generated anchor is assigned a lower weight $W$ in the new adaptive updating formula, denoted as $\overrightarrow{DP}^*$:
\highlight{
\begin{equation}
    \overrightarrow{DP}_{new}^*=
    \left\{ 
    \begin{array}{lcr} 
        \overrightarrow{DP}_{old}^* + \overrightarrow{N} \; &(IAI \leq IAI_0) \\
        \overrightarrow{DP}_{old}^* + W \cdot \overrightarrow{N} \; &(IAI > IAI_0)
    \end{array}\right.
    \label{Fupdate2}
\end{equation}
}
We set $W > 0$ to ensure that if the datapoint to be updated is empty, the new anchor can at least fill the blank and make it ready for query.
Based on experiments, we find that setting $k_{IAI}=0.95$, $IAI_0=1$, $W=0.1$ yields the best performance.
This adaptive updating scheme enhances the database accuracy, consequently improving orientation estimation.
\subsection{Distortion Detection}\label{determine use or not}
In scenarios where magnetic distortion is minimal, such as the conference room and outdoor environments as indicated in Table \ref{places}, the database offers little contribution to orientation estimation. 
Conversely, in distortion-free environments, deactivating the database can conserve computational load.
To identify distortion-free scenarios, we utilize the magnetic field magnitude $M$, which provides rich information on magnetic distortion, and can be conveniently measured using the magnetometer, via calculating the norm of its reading.
Using $M$, we design two criteria to detect the distortion-free places.
When both criteria are satisfied, we classify a place as distortion-free. 
If the place is detected as distortion-free (<10°), \aliasSystem automatically deactivates the database.

\begin{figure}[ht]
    \vspace{-2mm}
    \hfill
    \begin{minipage}[t]{0.45\linewidth}
        \centering
        \includegraphics[width=1\linewidth]{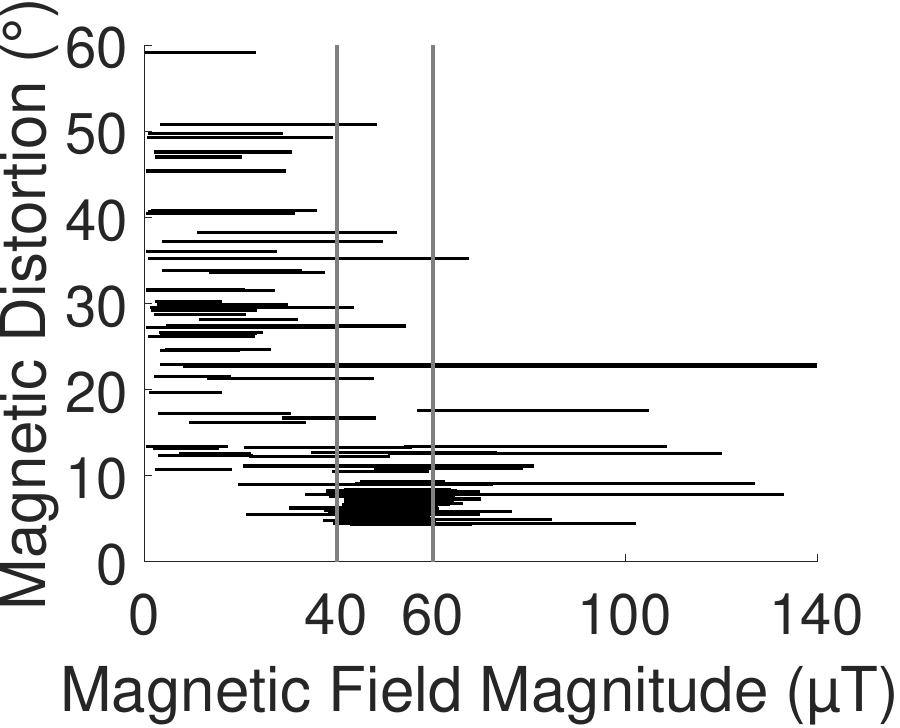}
         \vspace{-5mm}
        \caption{Relationship between distortion and magnetic field magnitude.}
        \label{Criterion_A}
    \end{minipage}
    \hfill
    \begin{minipage}[t]{0.46\linewidth}
        \centering
        \includegraphics[width=1\linewidth]{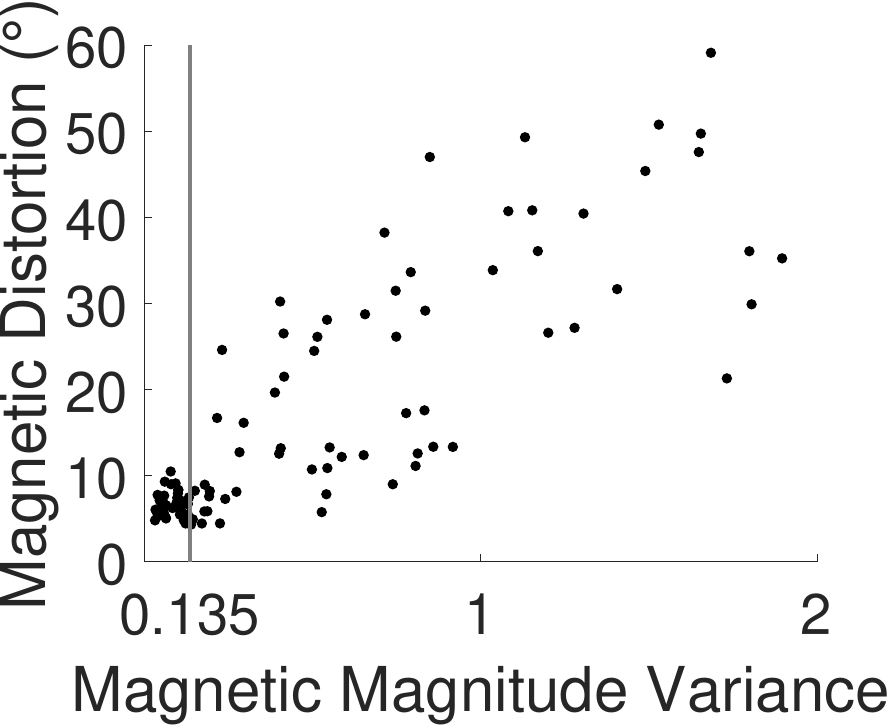}
         \vspace{-5mm}
        \caption{Relationship between magnetic distortion and magnitude variance.}
        \label{Criterion_B}
    \end{minipage}
    \hfill
    \vspace{-5mm}
\end{figure}

\begin{itemize} [leftmargin=*]
    \item \textbf{Criterion A:} 
    \textit{Range of $M$.}
    Distortion-free magnetic fields typically exhibit a magnitude distribution ranging from 45$\mu T$ to 55$\mu T$.
    Criterion A is considered true if all magnitude measurements fall within the range of 40$\mu T$ to 60$\mu T$.
    Figure \ref{Criterion_A} illustrates the distribution of magnetic field magnitude of all collected data traces, where each line represent the range of magnetic field magnitude (minimum on the left, maximum on the right).
    In instances where the magnitude range deviates from 40$\mu T$ to 60$\mu T$, distortion tends to be higher.
    There are also many lines that fall within 40$\mu T{\sim}60\mu T$, whose corresponding distortion is low (<10°).
    They overlap and may not be distinctly visible in Figure~\ref{Criterion_A}.

    \item \textbf{Criterion B:} 
    \textit{Variance of $M$.}
    Distortion in magnetic fields often correlates with the relative variance of magnetic field magnitude, expressed as $\frac{\sigma_M}{\overline{M}}$, where $\sigma_M$ is the deviation of $M$ and $\overline{M}$ is the average. 
    Figure \ref{Criterion_B} illustrates the distribution of magnetic field magnitude variances in all collected data. Higher variance typically indicates increased distortion in magnetic fields. 
    Criterion B is considered true if the relative variance in the area is lower than 0.135.
\end{itemize}




\vspace{-1mm}
\section{Evaluation} \label{sec_evaluation}
In this section, we evaluate the performance of \aliasSystem across various magnetic distortion levels, different locations, and among different users. 
We also quantify the contributions of different design components within \aliasSystem.

\subsection{Experiment Setting}
\label{secExperimentSetting}

Our main evaluation focuses on the orientation estimation error, defined as the minimum angle of rotation required to align the estimated orientation with the ground truth orientation. 
Throughout this work, \aliasSystem operates without access to any pre-scanned or pre-built database. 
Instead, \aliasSystem treats all scenarios as new, necessitating the construction of its database from scratch to model the local magnetic field.

\subsubsection{Baselines} \label{baselines}

    

    \textit{\textbf{MUSE}}\cite{muse}\textbf{:} The state-of-the-art arm tracking method, which uses complementary filter for orientation estimation and particle filter for location estimation. 
    
    \textit{\textbf{AVOID:}} 
    We also implement a custom orientation estimation baseline, $AVOID$, using the idea from work \cite{fan2017adaptive, ahrs}.
    It mitigates the negative impact of magnetic distortion by dynamically adjusting the weight of the magnetometer.
    We mainly implement the proposed strategy from the later one \cite{fan2017adaptive}, which anticipates the distortion intensity via observing the magnitude and the dip angle of the magnetic field. 
    \begin{equation}
        \begin{split}
            \lambda_1 &= \frac{(|M-M_0)|}{M_0} \:(if\: \lambda_1>1, let\: \lambda_1=1) 
            \\ 
            \hspace{9mm}
            \lambda_2 &= \frac{|\theta_{dip}-\theta_0|}{20\degree} \:(if\: \lambda_2>1, let\: \lambda_2=1) 
            \\ 
            \hspace{11mm}
            \lambda &=\frac{\lambda_1+\lambda_2}{2} \label{lambda}
        \end{split}
    \end{equation} 
As shown in Equation \ref{lambda}, it measures the magnetic field magnitude $M$ and compares it to the normal value of local geological magnetic field magnitude $M_0$ (usually 45$\mu$T $\sim$ 55$\mu$T). 
It calculates the relative difference between $M$ and $M_0$ as $\lambda_1$.
It then transforms the magnetometer direction into GRF using estimated orientation, and calculates the angle between the magnetometer direction and the surface: $\theta_{dip}$.
It further compares the current dip angle with the normal dip angle $\theta_{0}$ ($\theta_{0}$ mainly depends on longitude and latitude), and calculates the difference $\lambda_2$ using a threshold of 20°.
Finally, it averages $\lambda_1$ and $\lambda_2$ into $\lambda$ to represent the intensity of distortion.
A ratio of $(1-\lambda)$ controls the weight of magnetometer, so that it partly avoids the negative impact of magnetic distortion.
We implement the strategy of $(1-\lambda)$ in a complementary filter to control the magnetometer weight: $k_{m}(1-\lambda)$.

    



\begin{figure} [htbp] 
    \vspace{-3mm}
    \subfigure[Watch Acc.]{\includegraphics[width=0.3\linewidth]{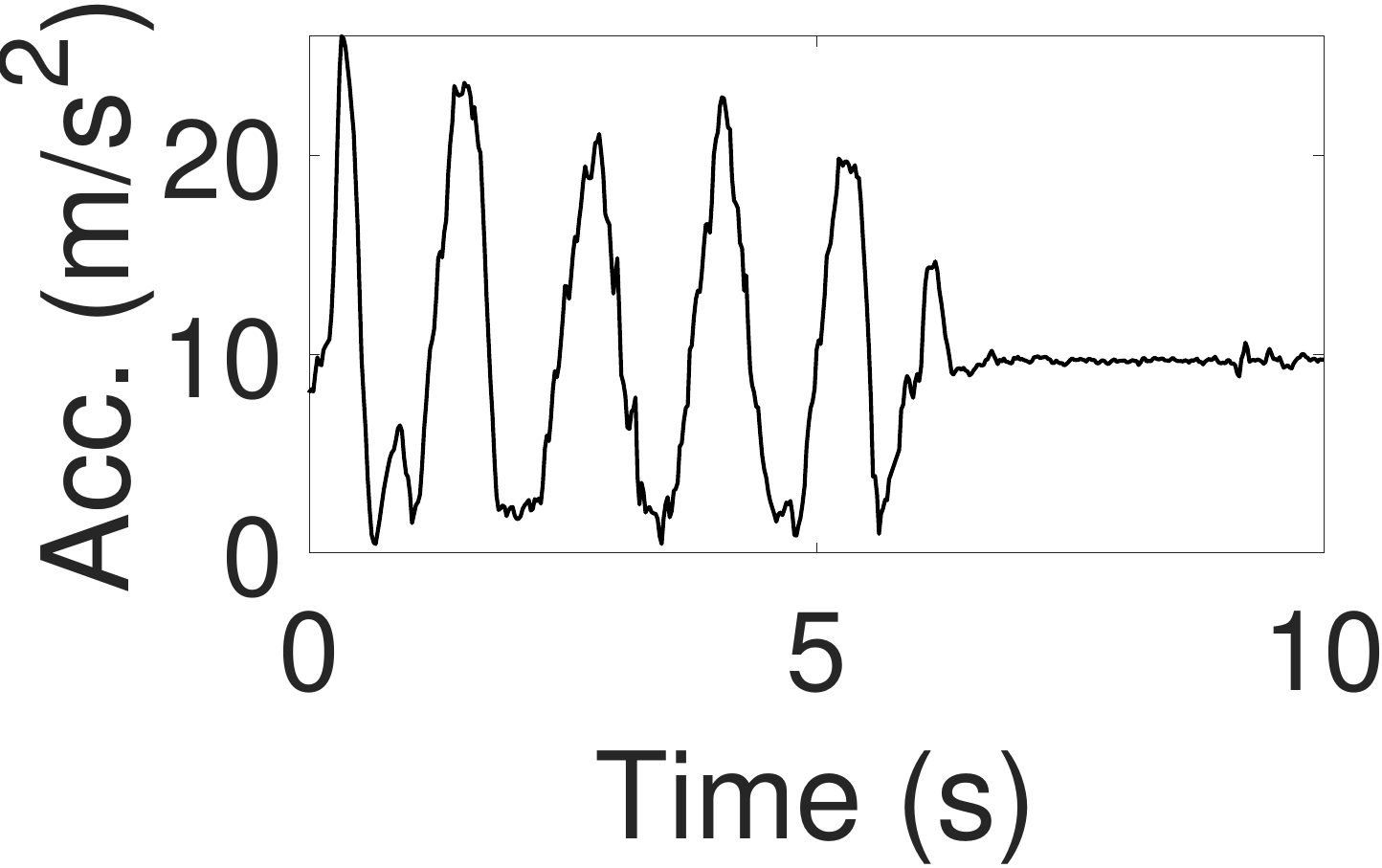}}
    \hfill
    \subfigure[Watch Gyro.]{\includegraphics[width=0.3\linewidth]{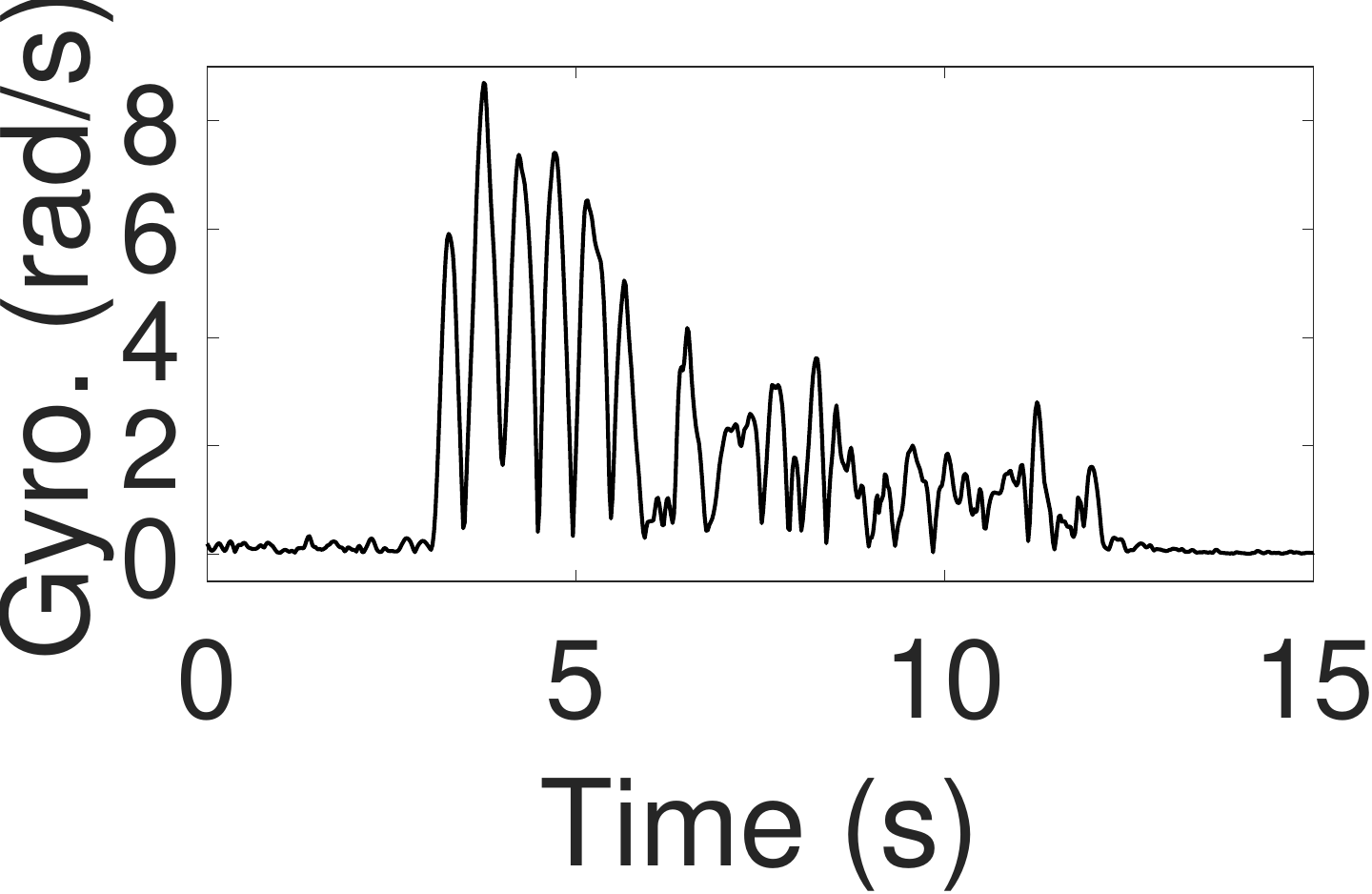}}
    \hfill
    \subfigure[Watch Mag.]{\includegraphics[width=0.3\linewidth]{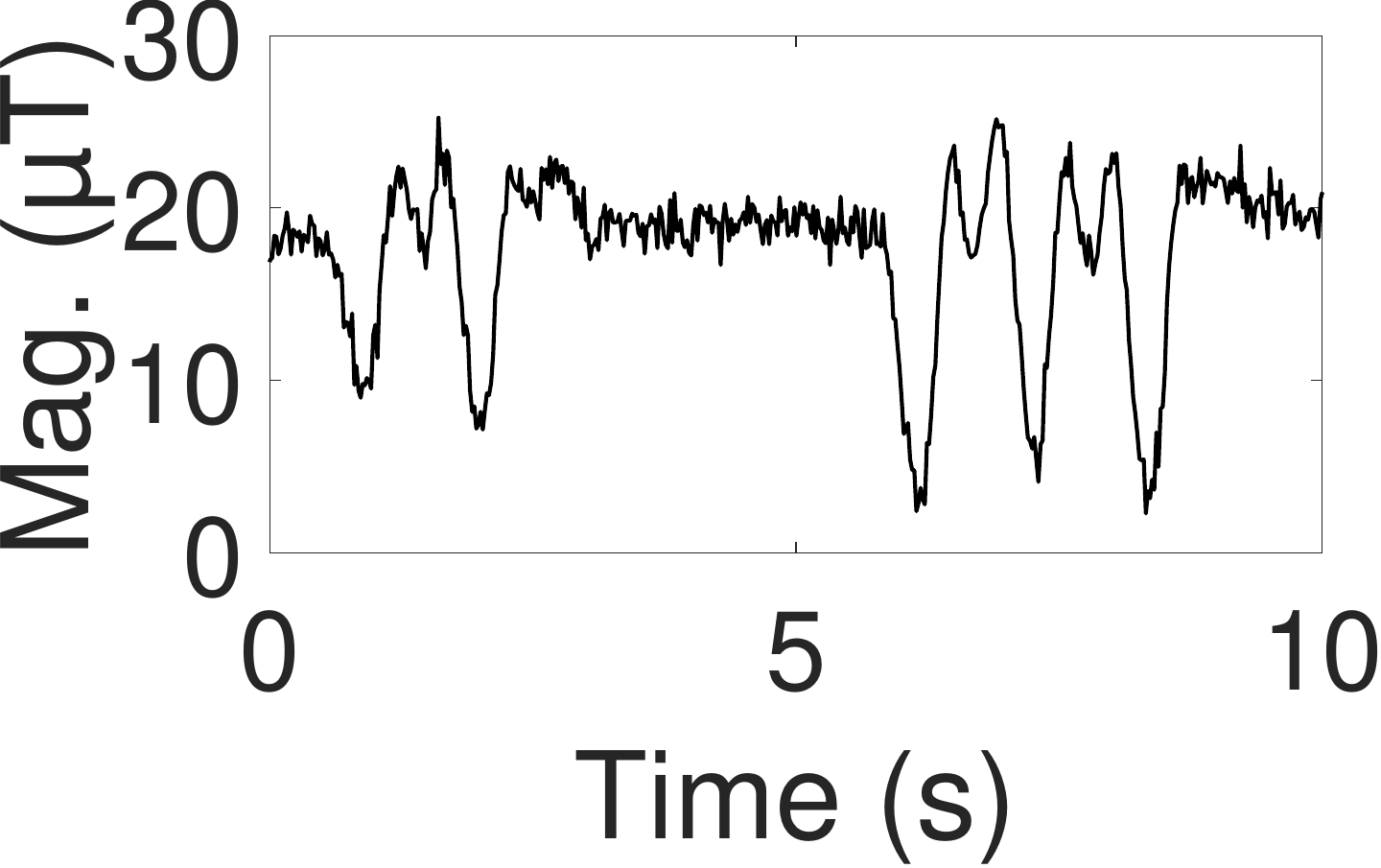}}
    
    \vspace{-3mm}
    \subfigure[Samsung Acc.]{\includegraphics[width=0.3\linewidth]{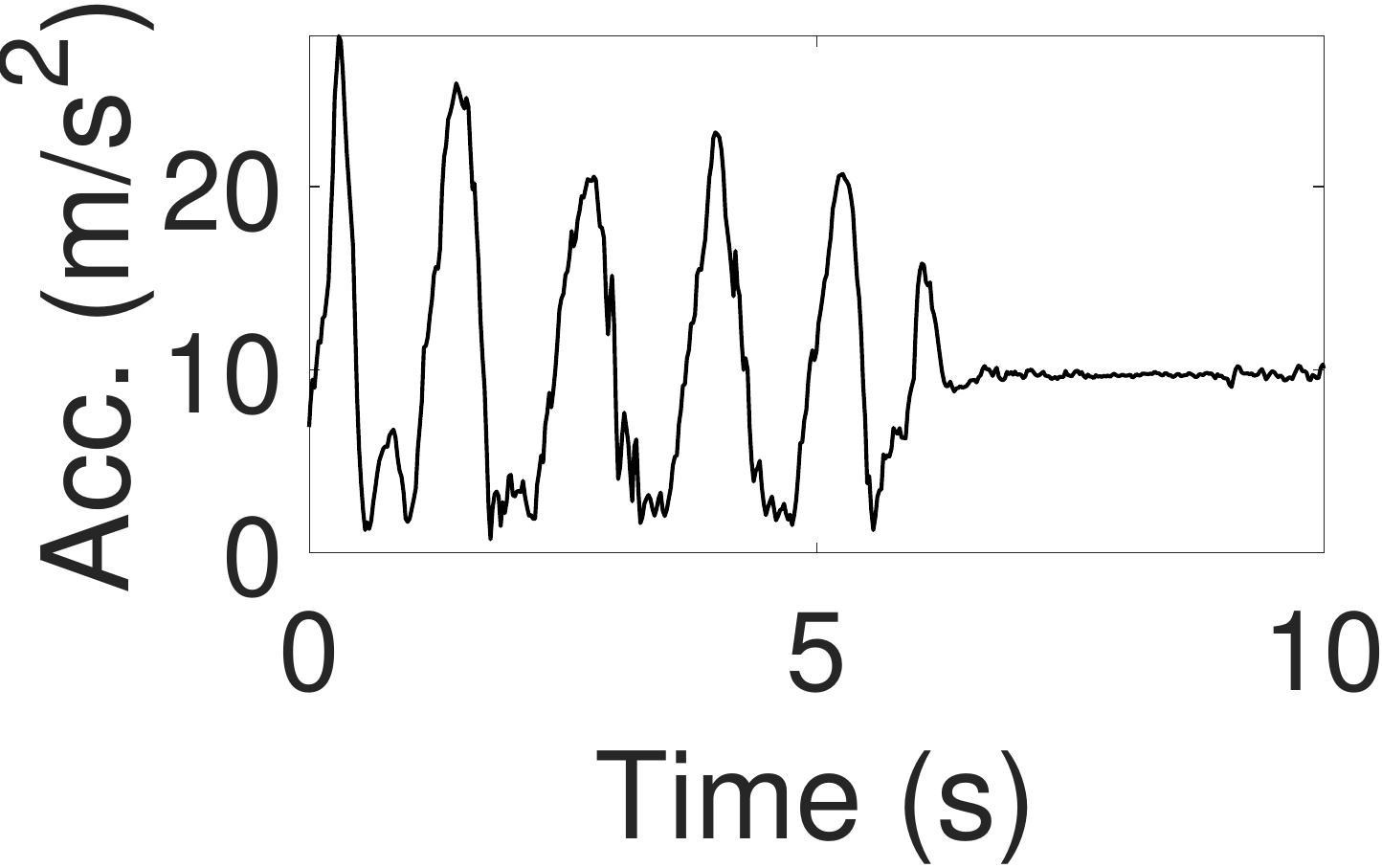}}
    \hfill
    \subfigure[Samsung Gyro.]{\includegraphics[width=0.3\linewidth]{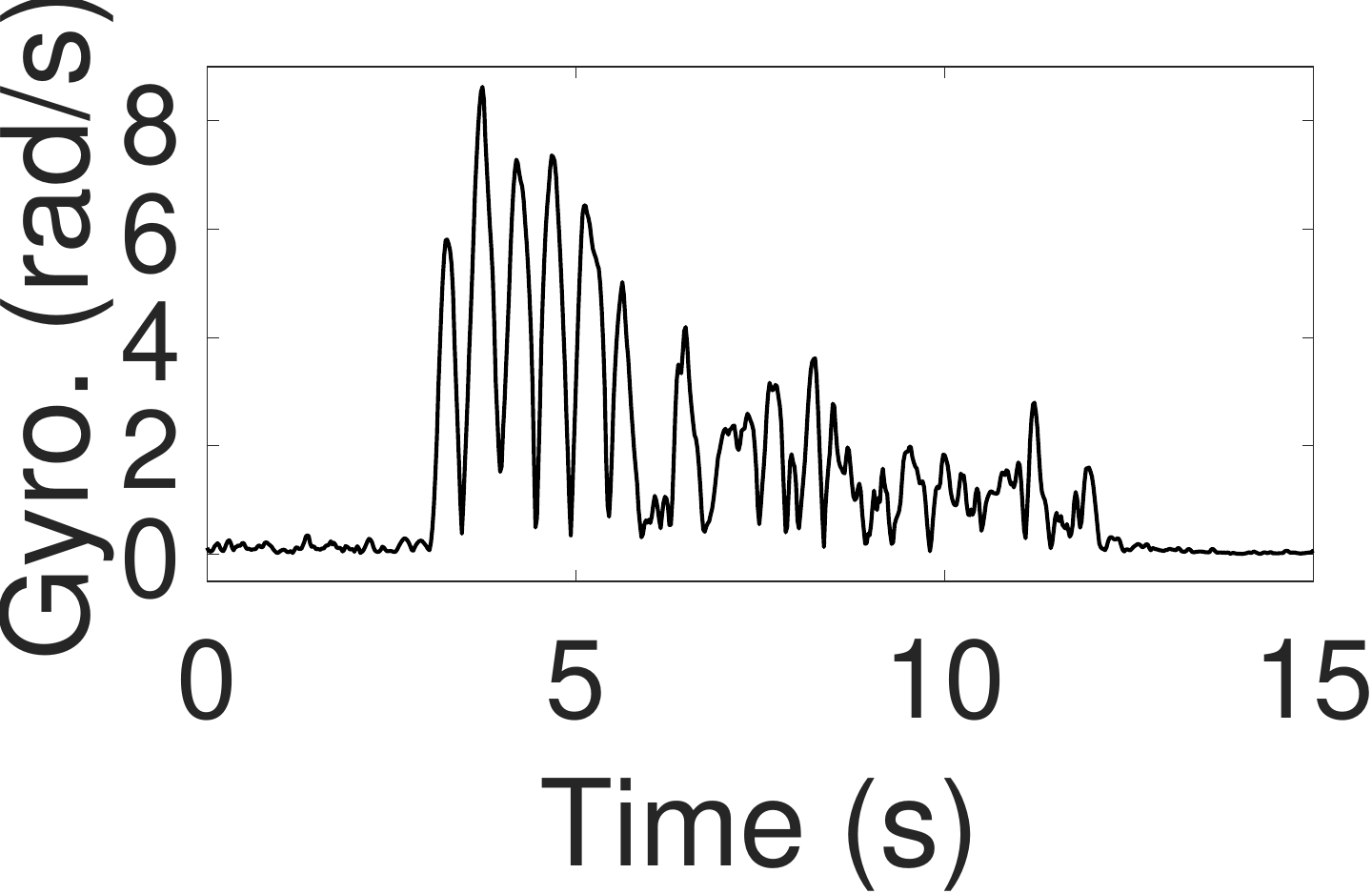}}
    \hfill
    \subfigure[Samsung Mag.]{\includegraphics[width=0.3\linewidth]{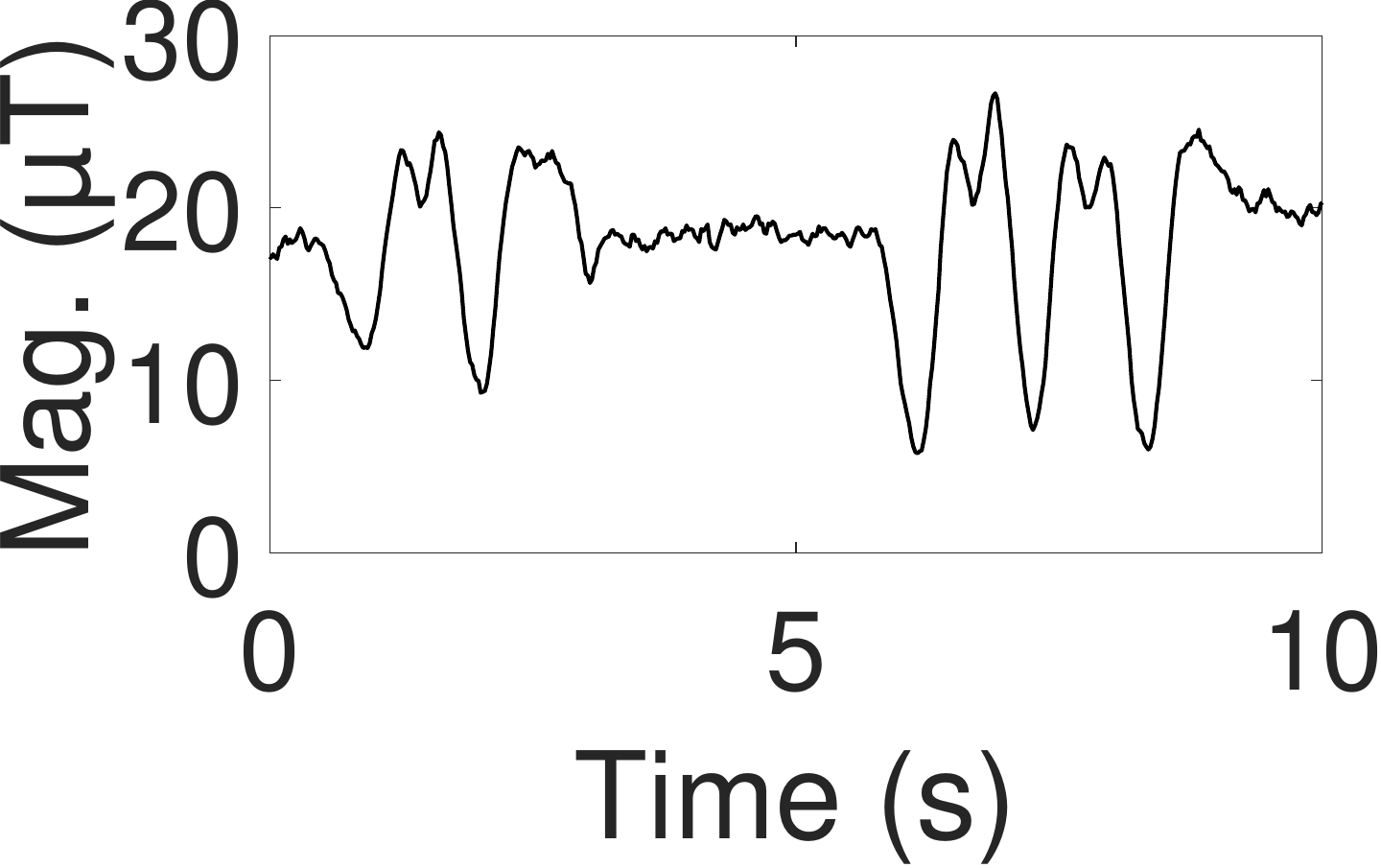}}
    
    \vspace{-3mm}
    \subfigure[iPhone Acc.]{\includegraphics[width=0.3\linewidth]{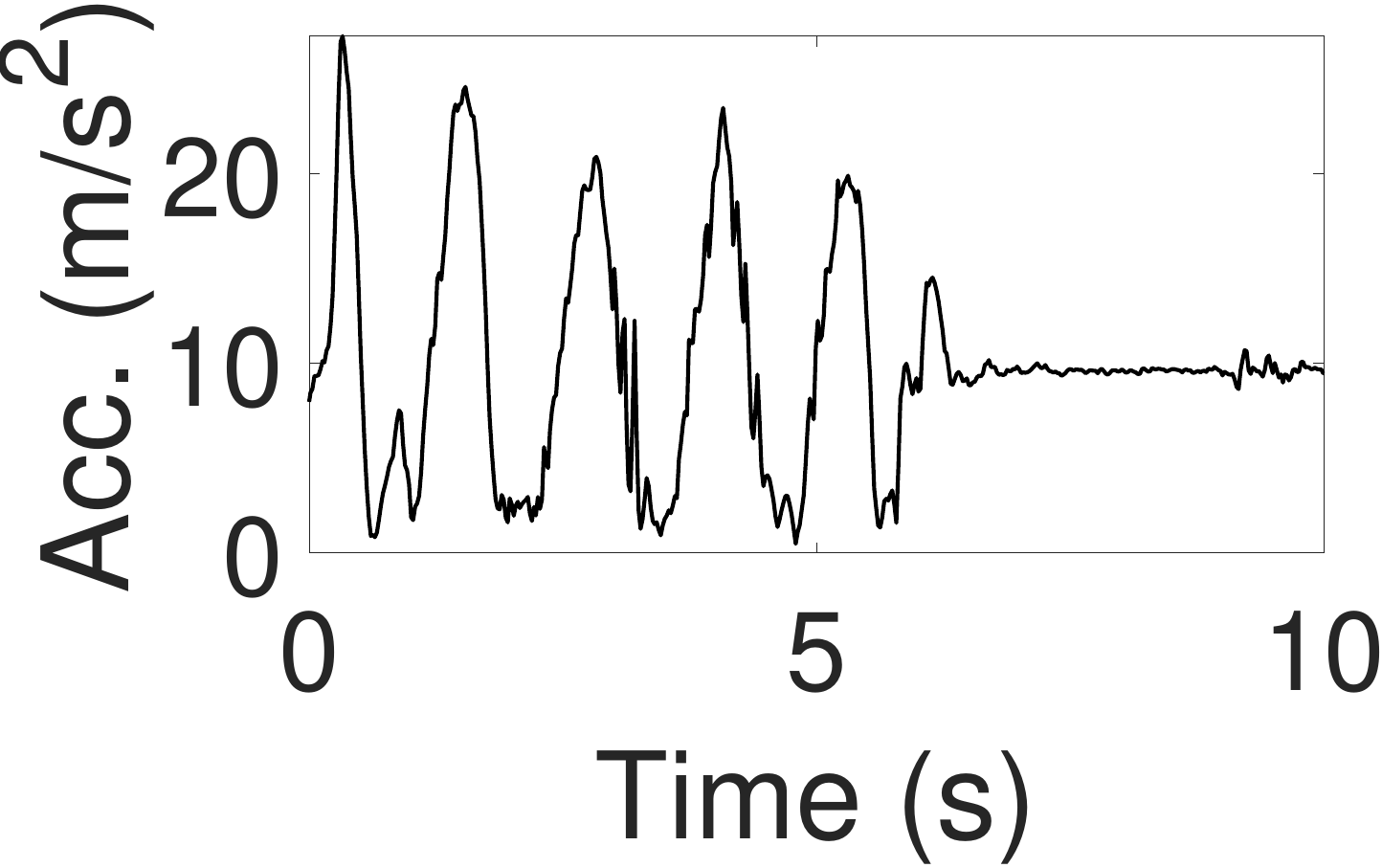}}
    \hfill
    \subfigure[iPhone Gyro.]{\includegraphics[width=0.3\linewidth]{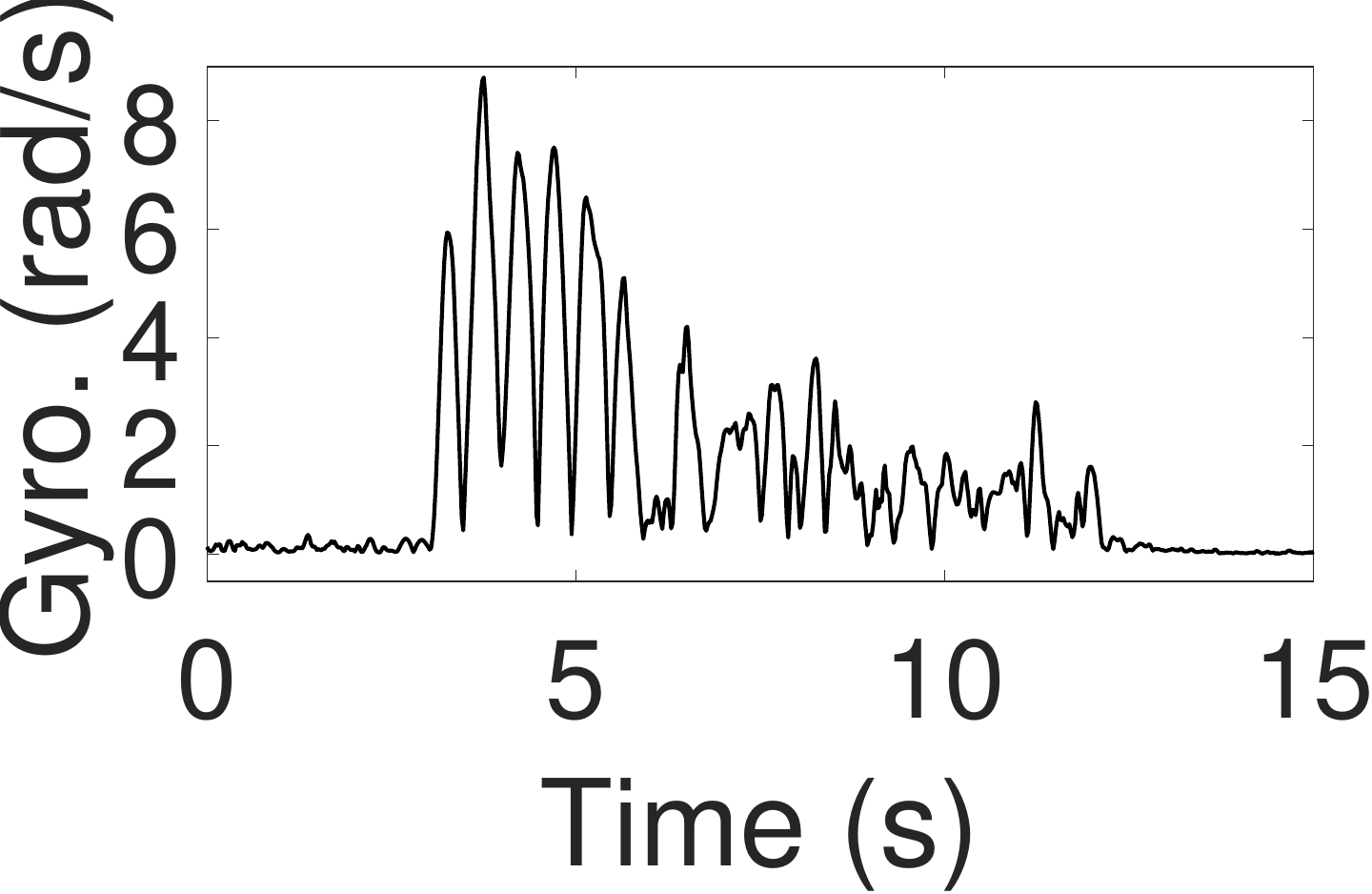}}
    \hfill
    \subfigure[iPhone Mag.]{\includegraphics[width=0.3\linewidth]{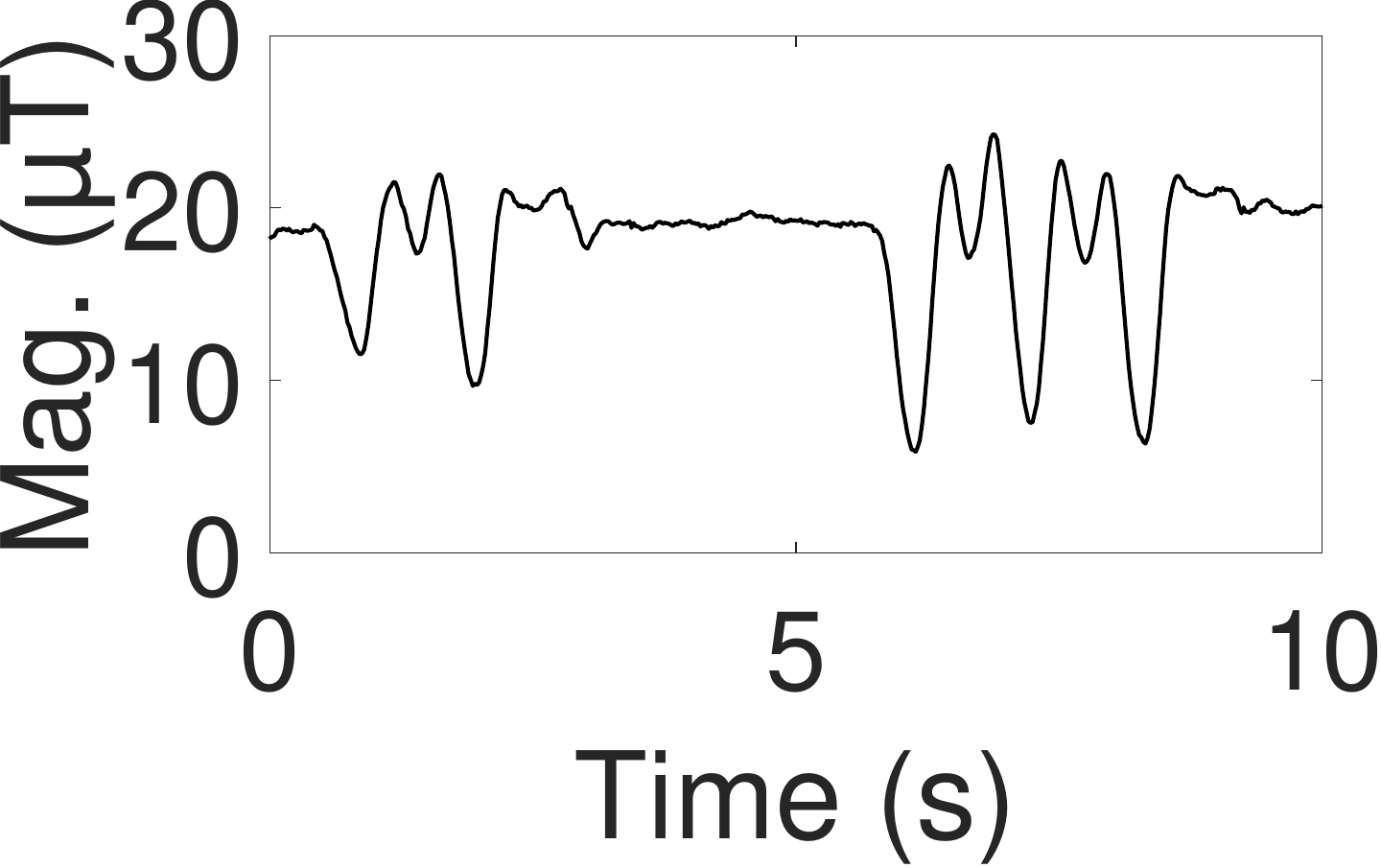}}
    
    \vspace{-2mm}
    
    \caption{Sensor comparison among smart devices}
    \label{compare}
    \vspace{-4mm}
\end{figure}

\subsubsection{Hardware} \label{hardware}
We use a Fossil Gen 5 smartwatch for real experiments. 
It is equipped with an AK0991X magnetometer and a LSM6DSO 6-axis accelerometer-gyroscope.
The sampling rate is 50 Hz for all three sensors.
Other mobile devices that use the same sensors include Sony Xperia 1 III, Oneplus 7, Motorola moto g fast, MiWatch, Xiaomi Mi 10T Pro, and TicWatch Pro 3.

To verify \aliasSystem's scalability to other platforms, we conduct real experiments to compare the smartwatch sensors with a Samsung Galaxy S21 5G and an iPhone XS max.
We bind them closely onto one solid plate and collect motion data in the corridor.
Figure \ref{compare} shows the magnitude of the sensors' raw readings from three smart devices.
Results indicate that IMU sensors on the smartwatch and the smartphones are similar.
After basic denoising, the sensors' relative similarities (1-$\frac{2|x-y|}{|x|+|y|}$) between the smartwatch and the Samsung Galaxy S21 are 91.91$\%$, 95.18$\%$, and 97.76$\%$, for accelerometer, gyroscope, and magnetometer respectively.
Similarities between the smartwatch and the iPhone are 97.25$\%$ 97.34$\%$ , and 98.16$\%$.
The similarity among the IMU sensors on different mobile platforms suggests that \aliasSystem can conveniently migrate to other platforms and maintain similar performance.

    

    

    
    

\subsubsection{Data Collection}
\label{data}

We invite 12 users to collect 100+ data traces, at 10 different places in 2 different cities, with a total duration of 27.53 hours.
Each data trace is around 10 $\sim$ 20 minutes. 
At the beginning of each data trace, the user stands still for 10 seconds, so that the system can initialize the GRF.
During the data collection, the user performs daily arm motion, such as:
\begin{itemize}
    \item Point to different directions, e.g., front, left, and right.
    \item Draw straight lines, e.g., up to down, and left to right.
    \item Write letters in air, from 'A' to 'Z'.
    \item Exercising, e.g., lifting dumbbells.
\end{itemize}
To better simulate daily arm movement, we allow the users to perform the above activities in any order they want.
Users sometimes perform customized arm motion, which is also recorded in the data traces.

\begin{table} [H]
    \vspace{-1mm}
    \small
    \centering 
        \caption{Data Statistics}
        \vspace{-2mm}
    \label{places}
    \begin{tabular}{|l|c|c|c|c|}
        \hline
        Places & City & 
        \begin{tabular}[c]{@{}c@{}}
            Distortion 
        \end{tabular}  & 
        Duration 
        & 
        \begin{tabular}[c]{@{}c@{}}
            D.B. 
            Improv. 
        \end{tabular} \\
        \hline
        Conf. room & $\#$1 & 6.93° & 10.8 hours  
        & -1.83$\%$  \\
        \hline
        Outdoor & $\#$1 & 6.63° & 58 minutes 
        & 3.43$\%$ \\
        \hline
        House & $\#$1 & 7.06° & 48 minutes 
        & 4.37$\%$   \\
        \hline
        Apartment $\#$2 & $\#$1 & 11.65° & 35 minutes 
        & 9.20$\%$ \\
        \hline
        Stairwell & $\#$1 & 14.12° & 48 minutes 
        & 23.47$\%$ \\
        \hline
        Office & $\#$1 & 14.51° & 48 minutes 
        & 27.80$\%$ \\
        \hline
        Apartment $\#$3 & $\#$2 & 16.46° & 34 minutes 
        & 33.59$\%$ \\
        \hline
        Apartment $\#$1 & $\#$1 & 17.05° & 38 minutes 
        & 25.08$\%$  \\
        \hline
        Auditorium & $\#$1 & 28.51° & 35 minutes 
        & 23.91$\%$ \\
        \hline
        Corridor & $\#$1 & 31.06° & 11 hours 
        & 34.66$\%$ \\
        \hline
    \end{tabular}
    \vspace{-3mm}
\end{table}

\begin{figure} [ht]
    \subfigure[Conference Room.]{\includegraphics[width=.15\textwidth]{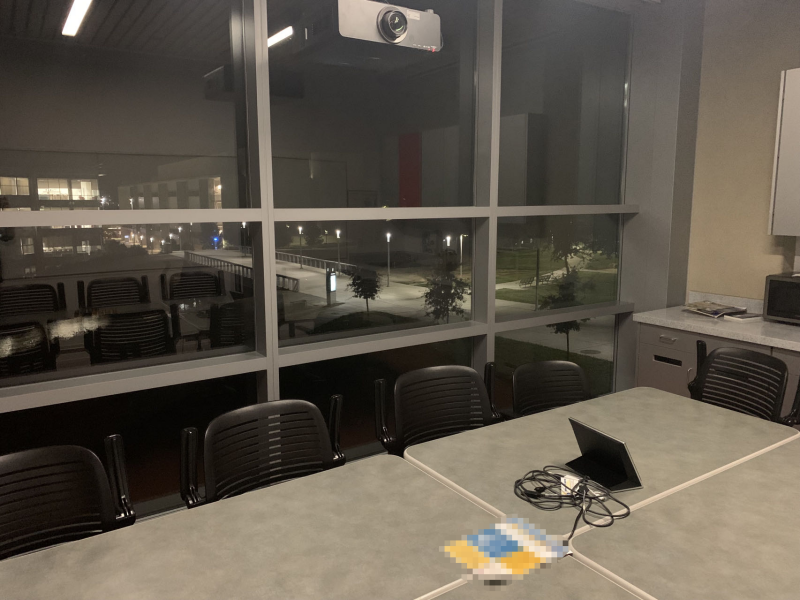}}
    \hfill
    \subfigure[Corridor.]{\includegraphics[width=.15\textwidth]{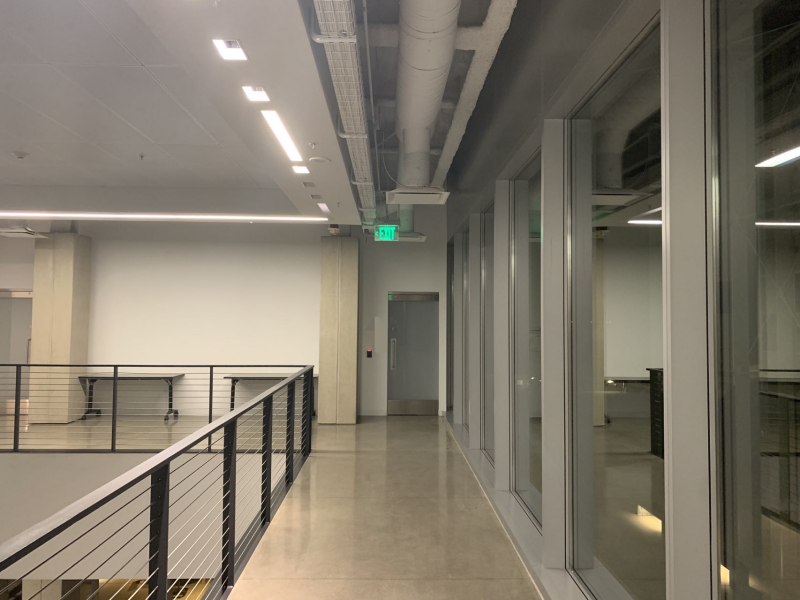}}
    \hfill
    \subfigure[House.]{\includegraphics[width=.15\textwidth]{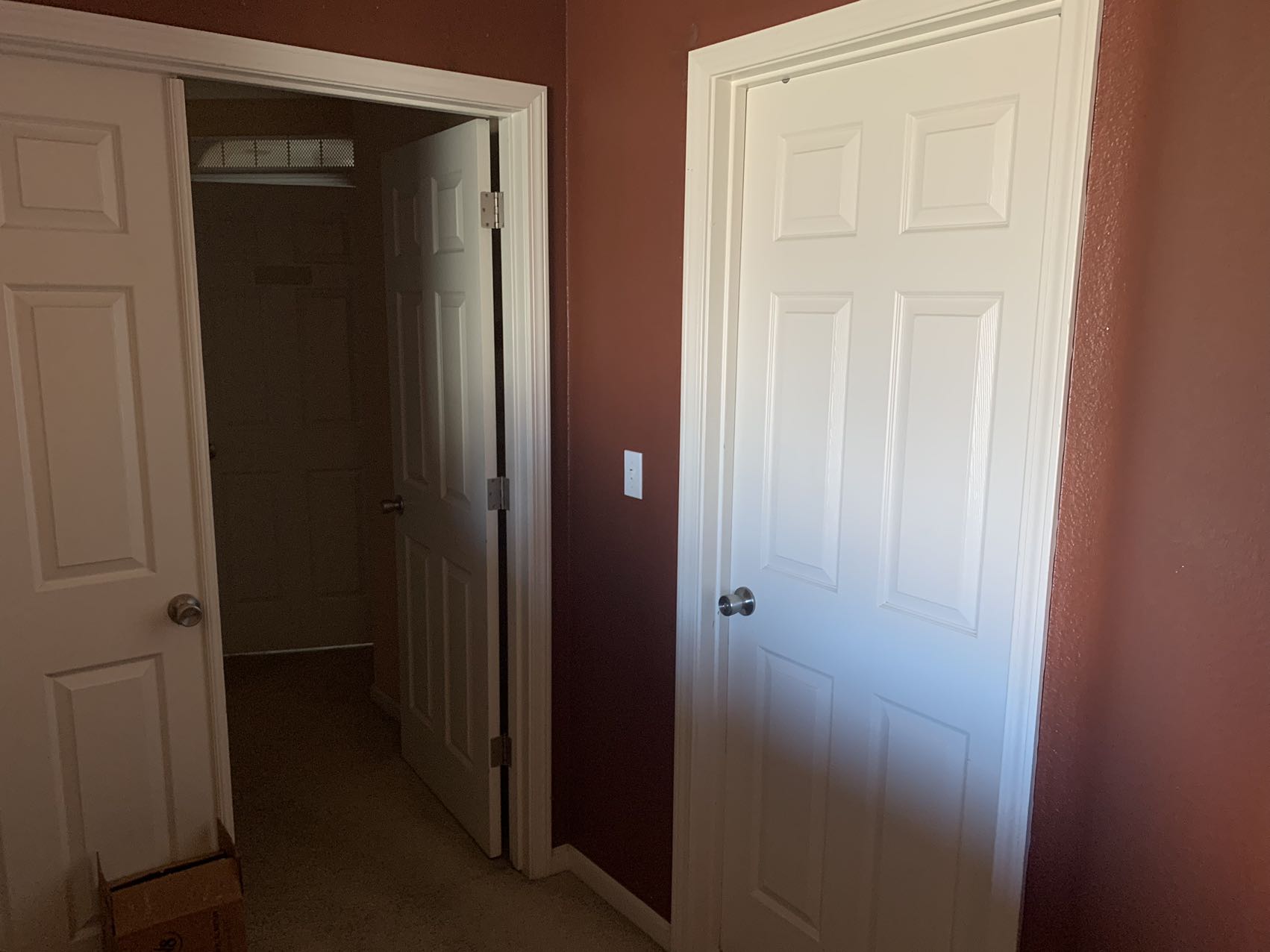}}

    \vspace{-3mm}
    
    \subfigure[Stairwell.]{\includegraphics[width=.15\textwidth]{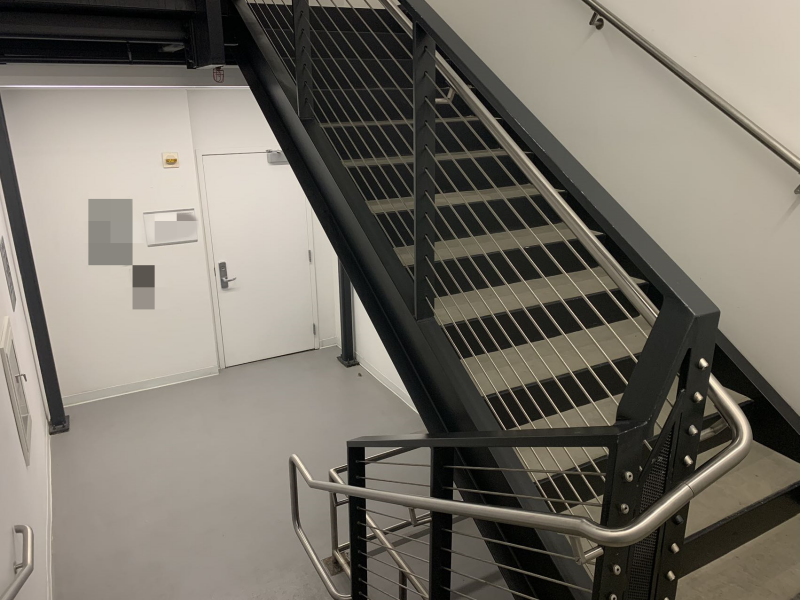}}
    \hfill
    \subfigure[Office.]{\includegraphics[width=.15\textwidth]{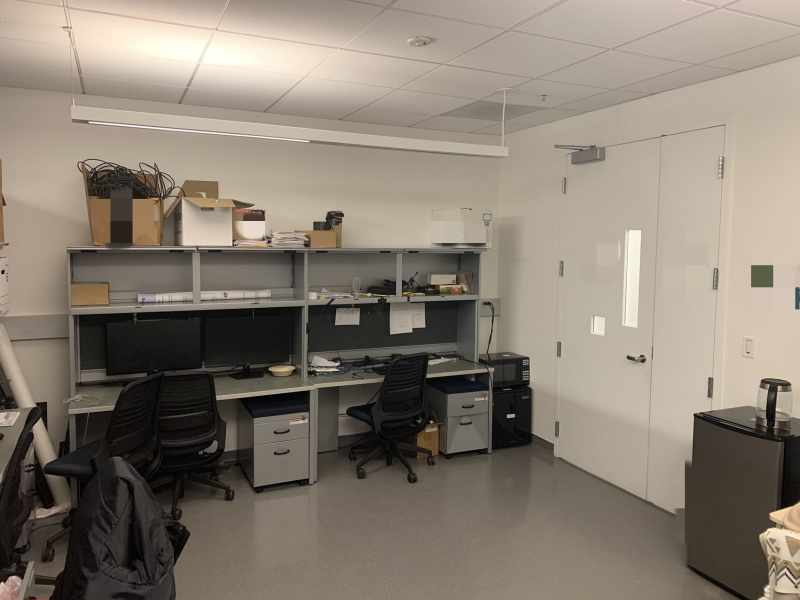}}
    \hfill
    \subfigure[Auditorium.]{\includegraphics[width=.15\textwidth]{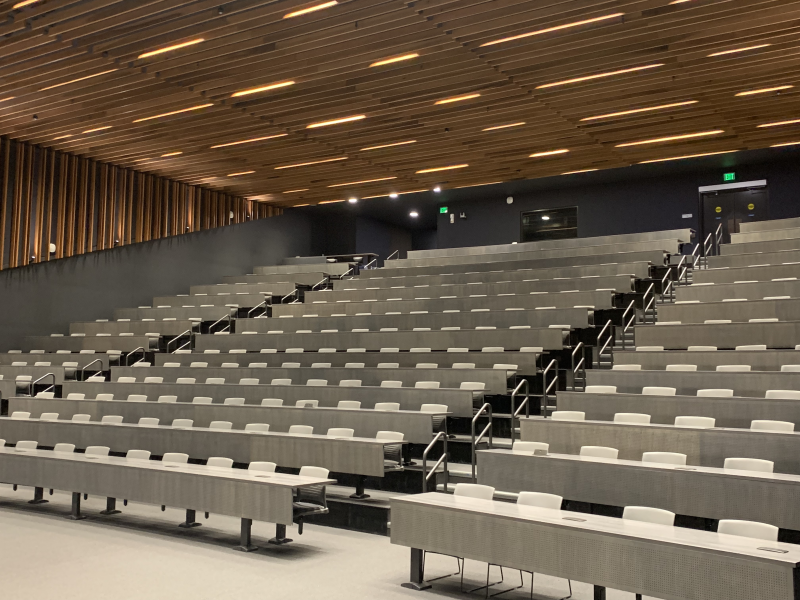}}

    \vspace{-3mm}

    \subfigure[Apartment 1.]{\includegraphics[width=.15\textwidth]{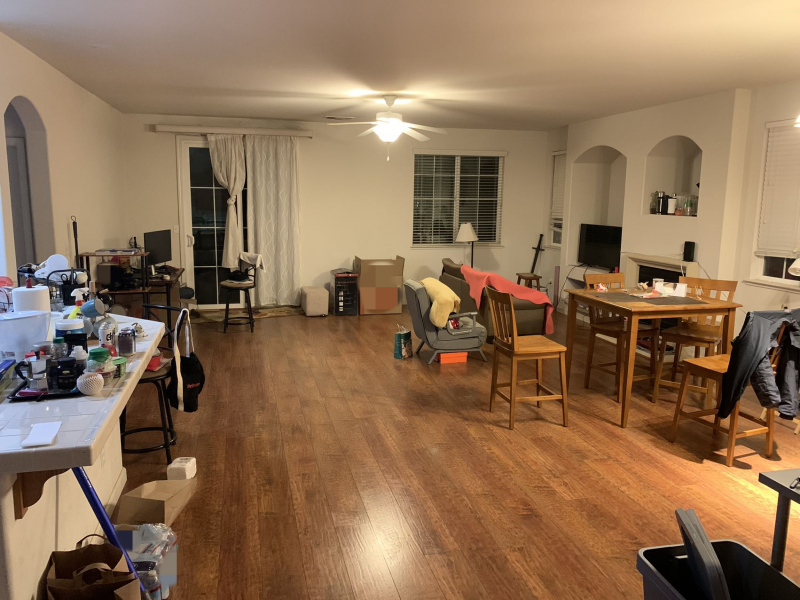}}
    \hfill
    \subfigure[Apartment 2.]{\includegraphics[width=.15\textwidth]{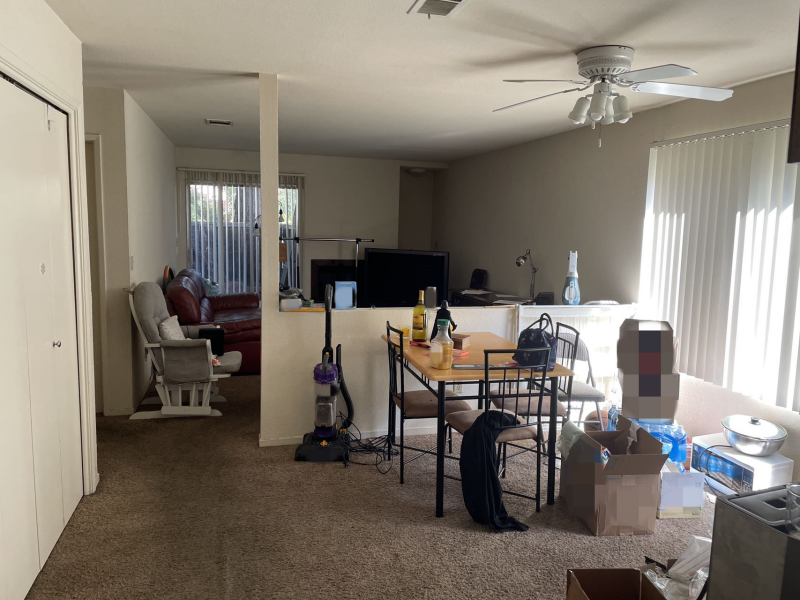}}
    \hfill
    \subfigure[Apartment 3.]{\includegraphics[width=.15\textwidth]{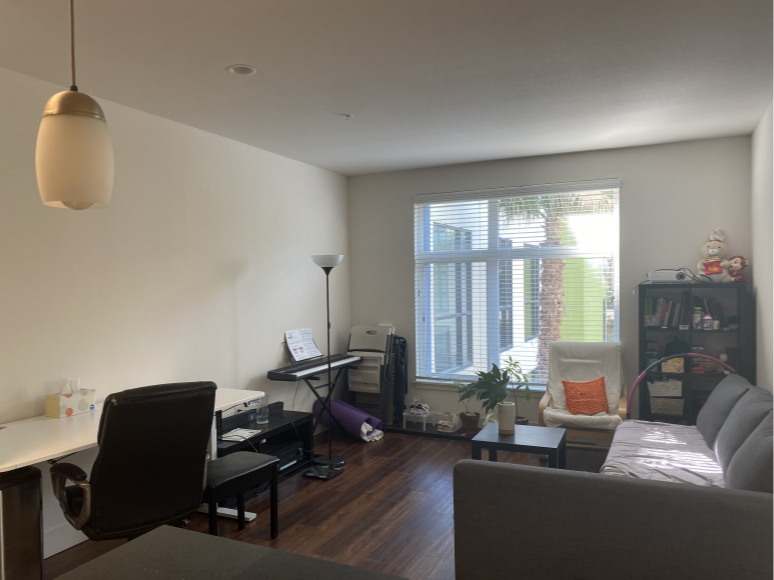}}
    \vspace{-2mm}
    \caption{Places for data collection and experiments}
    \label{fig_photo}
    \vspace{-1mm}
\end{figure}

We collect data in 10 typical places.
Table \ref{places} lists the places and their level of magnetic distortion.
The majority of the data are collected in the conference room (6.93°) and the corridor (31.06°), which are two representative places for light and heavy magnetic distortion.
Figure~\ref{fig_photo} (a) and (b) show their photos.
The conference room is separated from the outside by only a glass, resulting in light distortion.
In contrast, the corridor is surrounded by metal structure, leading to heavy distortion.
Geological position affects the pattern of magnetic field, with latitude having a greater influence than longitude.
Therefore we also collect data in the Apartment $\#$3 in City $\#$2.
City $\#$1 and City $\#$2 have a longitude difference of 2° and a latitude difference of 10°.

\subsubsection{Ground Truth}
\label{ground truth}
Ground truth was collected using the technique developed in \cite{miao}.
It uses a Meta Quest 2 to collect location and orientation ground truth, which adopts computer vision based tracking technique to locate the controller device, as well as sensing the orientation of the controller.
With the smartwatch solidly attached to the controller, its location and orientation can be acquired from the location and orientation of the controller through a constant transformation.
According to \cite{miao,rojo2022accuracy, holzwarth2021comparing, jost2021quantitative}, the Meta Quest 2 exhibits an orientation error below 0.85° and a location error below 0.7~cm.



\subsection{Overall Evaluation} \label{eval_overall}
\begin{figure*}[ht]
\centering
    \vspace{-1mm}
    \hfill
    \begin{minipage}[t]{0.4\linewidth}
        \centering
        \includegraphics[width=1\linewidth]{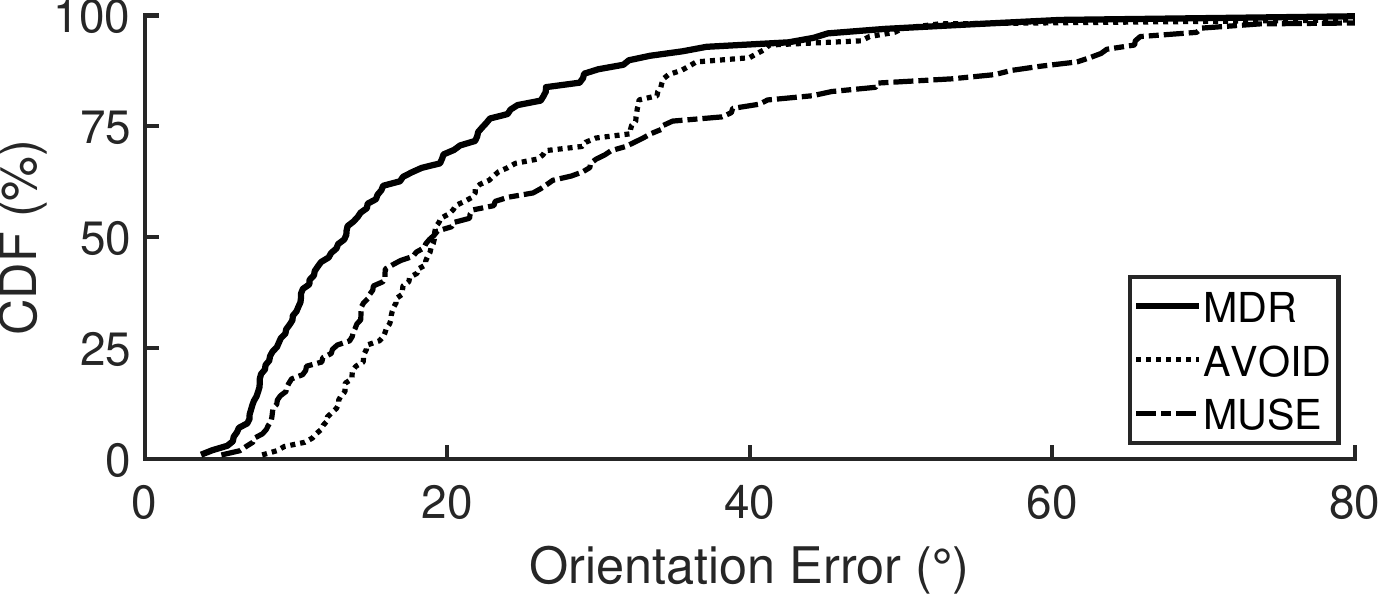}
        \vspace{-5mm}
       \caption{Overall orientation error.}
        \label{overall}
    \end{minipage}
    \hfill
    \begin{minipage}[t]{0.46\linewidth}
        \centering
        \includegraphics[width=1\linewidth]{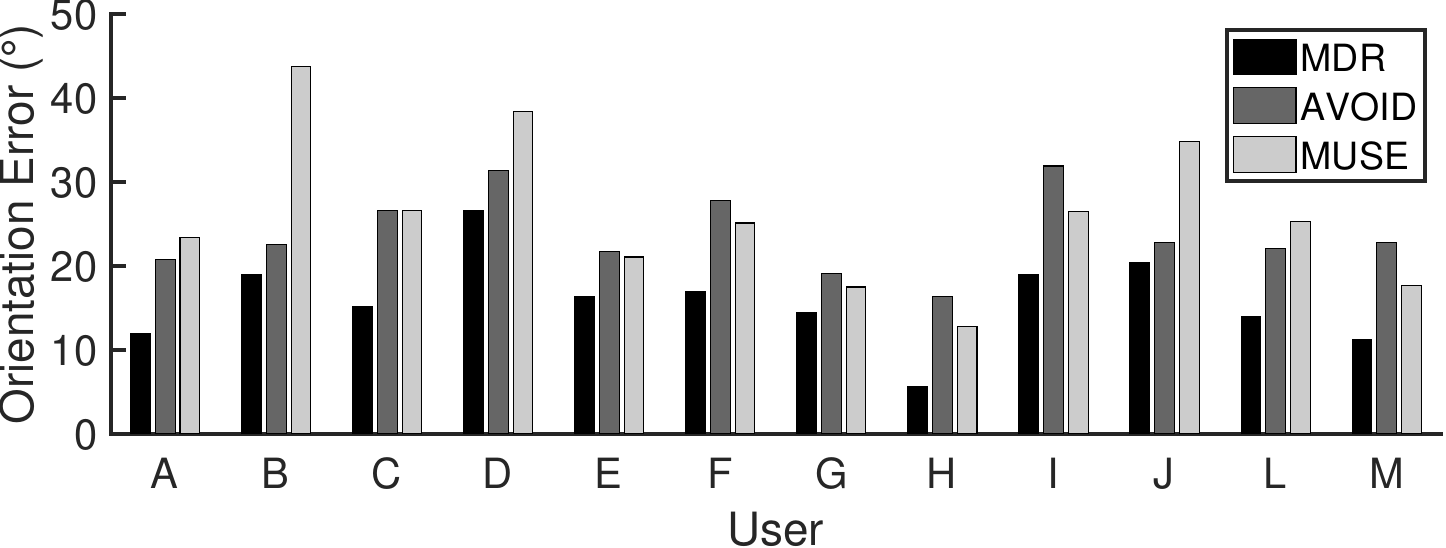}
        \vspace{-5mm}
       \caption{Performance among different users.}
        \label{users}
    \end{minipage}
    \hfill
    \vspace{-3mm}
\end{figure*}

Figure \ref{overall} shows the overall orientation estimation performance.
The overall average orientation errors are 17.73°, 22.96°, and 27.42° for \aliasSystem, $AVOID$, and MUSE, respectively.
\aliasSystem has an overall improvement of 35.34$\%$ over MUSE and 22.80$\%$ over $AVOID$.
The performance gain over MUSE stems from the distortion resistance enabled via the database design, whereas the performance gain over $AVOID$ is attributed to the different approach in handling distortion.
To further analyze the performance of these methods, we also present the experiment results under different scenarios.

\subsubsection{Different Users} \label{eval_users}
Figure \ref{users} shows the average orientation error across different users' data. 
\aliasSystem outperforms both baselines among all users.
Different users may have different error, depending on the user's motion pattern, which mainly affects the accelerometer accuracy.
For example, user D moves arm quickly, resulting in an average accelerometer direction error of 16.72°, while user H moves slowly with an average accelerometer direction error of 7.79°.
Consequently, the overall orientation error on user D is higher than user H.




\vspace{-1mm}
\subsubsection{Different Places}\label{pla}
\begin{figure*} [ht]
    \vspace{-1mm}
    \hfill
    \begin{minipage}[t]{0.48\linewidth}
        \centering
        \includegraphics[width=.88\linewidth]{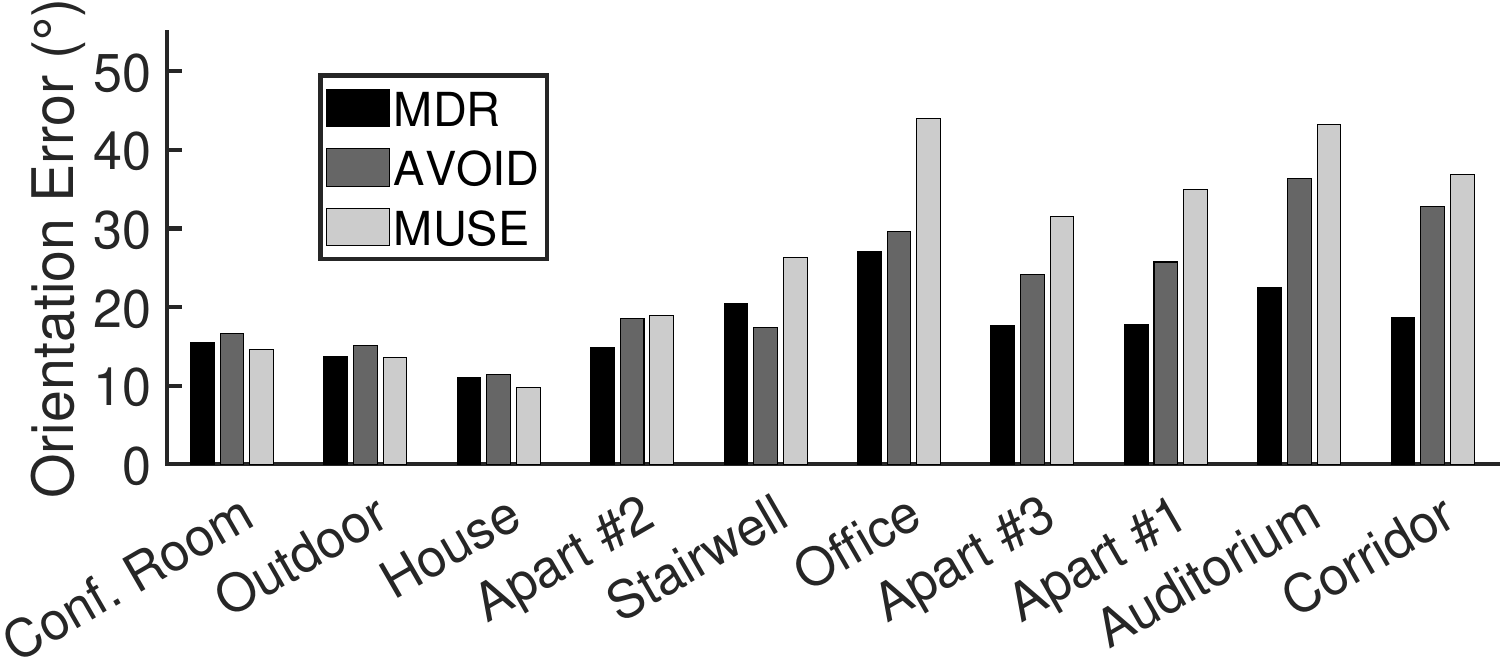}
        \vspace{-2mm}
       \caption{Performance at different places.}
        \label{place}
    \end{minipage}
    \hfill
    \begin{minipage}[t]{0.38\linewidth}
        \centering
        \small
        \includegraphics[width=1\linewidth]{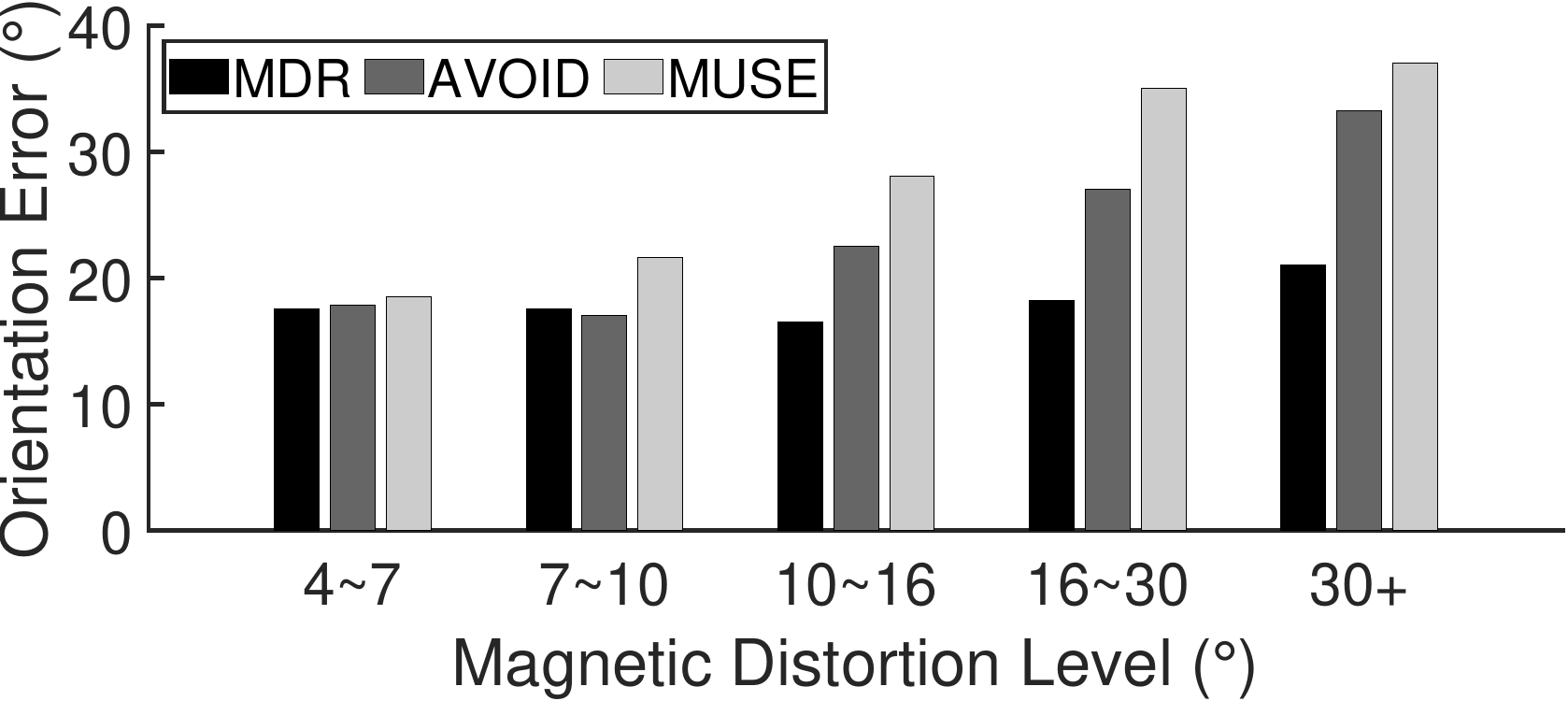}
        \vspace{-6mm}
        \caption{\small Different distortion levels.}
        \label{result}
    \end{minipage}
    \hfill
    \vspace{-3mm}
\end{figure*}

Table \ref{places} summarizes the data collected at different places.
Figure \ref{place} shows the orientation errors at different places, which are sorted mainly according to their distortion level.
In most cases, \aliasSystem outperforms both baselines.
Especially for 'Corridor', whose distortion is the highest, \aliasSystem has an error of 18.73°, which is 39.44$\%$ lower than $AVOID$.
However, for 'Stairwell', \aliasSystem is worse than $AVOID$, possibly due to the nearby metal door that frequently opens and closes, which changes the magnetic distortion pattern.
Dynamic magnetic distortion pattern is unfriendly to the database, as it breaks the chronological stability assumption.
Therefore, in such rare case, $AVOID$, who simply avoids distortion, demonstrates better performance.

\vspace{-1mm}
\subsubsection{Different Magnetic Distortion Levels} \label{eval_distortion}
To further evaluate the distortion resistance of \aliasSystem, we analyze the distortion magnitude measured in each data trace. 
Figure \ref{result} illustrates the orientation error of the data traces, sorted and categorized by their distortion level.
As the distortion increases, MUSE's performance degrades, since it consistently relies on the magnetometer, which is susceptible to distortion.
$AVOID$ demonstrates relatively stable performance, confirming the anti-distortion capability of the strategy proposed by \cite{fan2017adaptive}.
However, $AVOID$ fails when distortion eventually reaches 30°.
We monitor the relative weight of the magnetometer ($\lambda$ in Equation \ref{lambda}) in $AVOID$ when distortion is around 30° and find average $\lambda$ is around 0.5.
This indicates that when distortion is high, $AVOID$ suffers both from distortion and the absence of magnetic calibration, as it only partially avoids using the magnetometer.
This highlights that merely avoiding distortion does not fundamentally solve the distortion problem in orientation estimation.
In contrast, \aliasSystem appears to be more stable as distortion level increases, verifying its effective magnetic distortion resistance capability of \aliasSystem.

\subsection{Performance Decomposition of \aliasSystem}\label{decompose}

\subsubsection{Effect of Database at Different Places}
Since magnetic distortion pattern can be different across places, we evaluate the contribution of database at different places.
Table \ref{places} shows the distortion levels at each places, as well as the relative improvement of database, which is examined by turning on/off database.
As distortion rises, the improvement via database also increases, verifying the contribution enabled by the database in resisting magnetic distortion.
For the most distorted 'Corridor' that has 31.06° of distortion, the relative contribution of database is 34.66$\%$.
However, in 'Conference Room', the relative improvement of database is -1.83$\%$.
This negative value occurs because the 'Conference Room' has little distortion, thus database will not improve orientation accuracy, while errors in the database may result in a slight decrease in orientation accuracy.
\highlight{This negative effect can be avoided by the distortion detection module introduced in Section \ref{determine use or not}, which will be evaluated in Section \ref{detection_eval}.}


\subsubsection{Effect of Adaptive Updating}\label{eval_adaptive_updating}
To evaluate the effect of the adaptive updating scheme, we compare \aliasSystem with a sub-baseline of \aliasSystem without adaptive updating (denoted as \textbf{\aliasSystem w/o AU}), which naively aggregates the anchors into datapoints, as shown in Equation \ref{Fupdate}.
We conducted complete data traces using both approaches.
After that, we record the final database error, which is the average direction error of all anchors stored in the database.
Average final database error of \aliasSystem is 13.24°, which is 9.07$\%$ lower than the 14.56° of \textbf{\aliasSystem w/o AU}.
This difference verifies the capability of the adaptive updating scheme to enhance the accuracy of the database.

\begin{figure*}[ht]
    \hfill
    \begin{minipage}[t]{0.23\linewidth}
        \centering
        \includegraphics[width=1\linewidth]{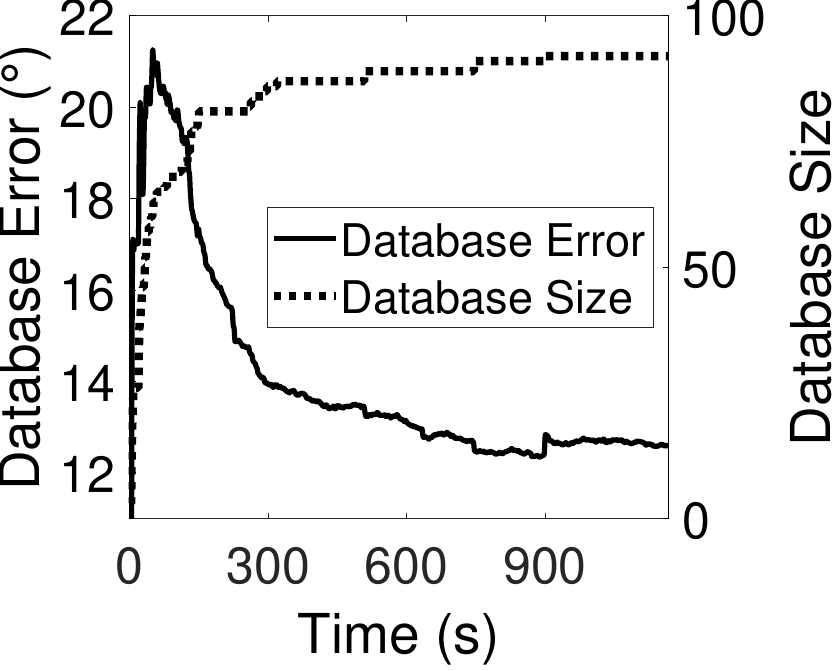}
        \vspace{-5mm}
        \caption{Database accuracy and database size during self-construction.}
        \label{build up}
    \end{minipage}
    \hfill
    \begin{minipage}[t]{0.23\linewidth}
        \centering
        \includegraphics[width=1\linewidth]{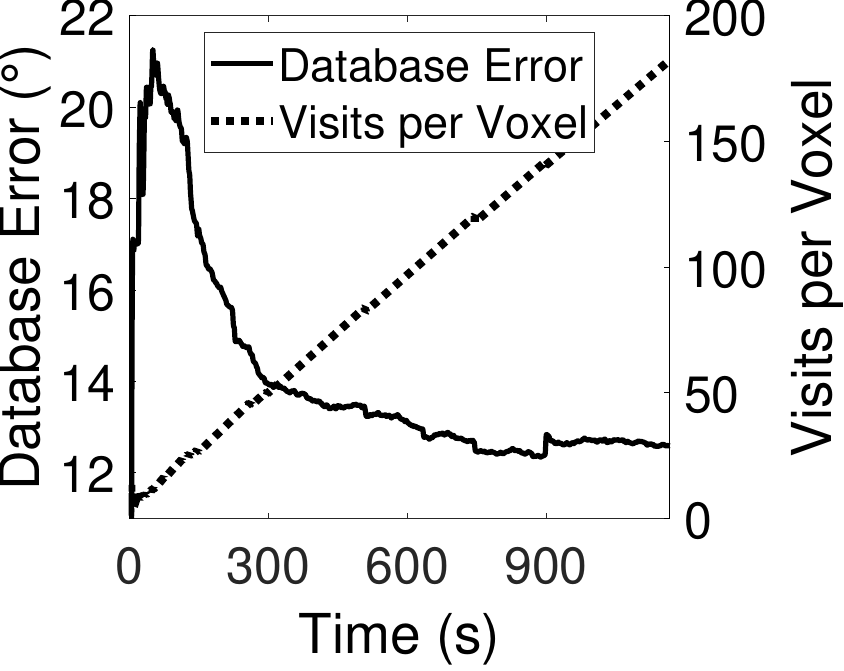}
        \vspace{-5mm}
        \caption{Database accuracy and visits per voxel during self-construction.}
        \label{build up 2}
    \end{minipage}
    \hfill
    \begin{minipage}[t]{0.18\linewidth}
        \centering
        \includegraphics[width=1\linewidth]{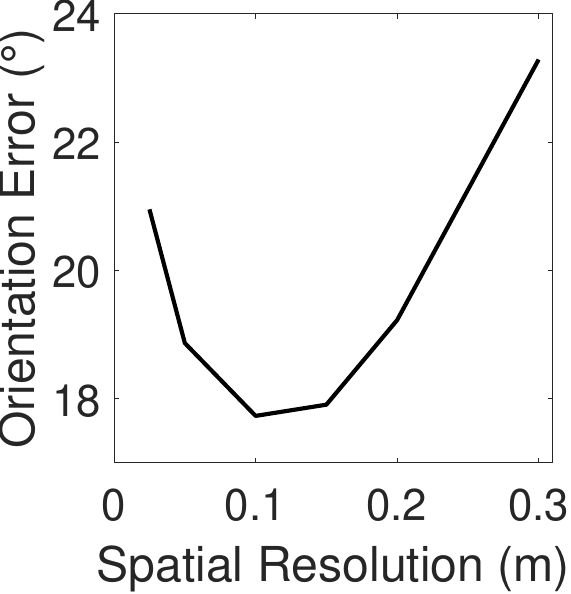}
        \vspace{-5mm}
        \caption{\aliasSystem error with different spatial resolution.}
        \label{prec}
    \end{minipage}
    \hfill
    \begin{minipage}[t]{0.21\linewidth}
        \centering
        \includegraphics[width=1\linewidth]{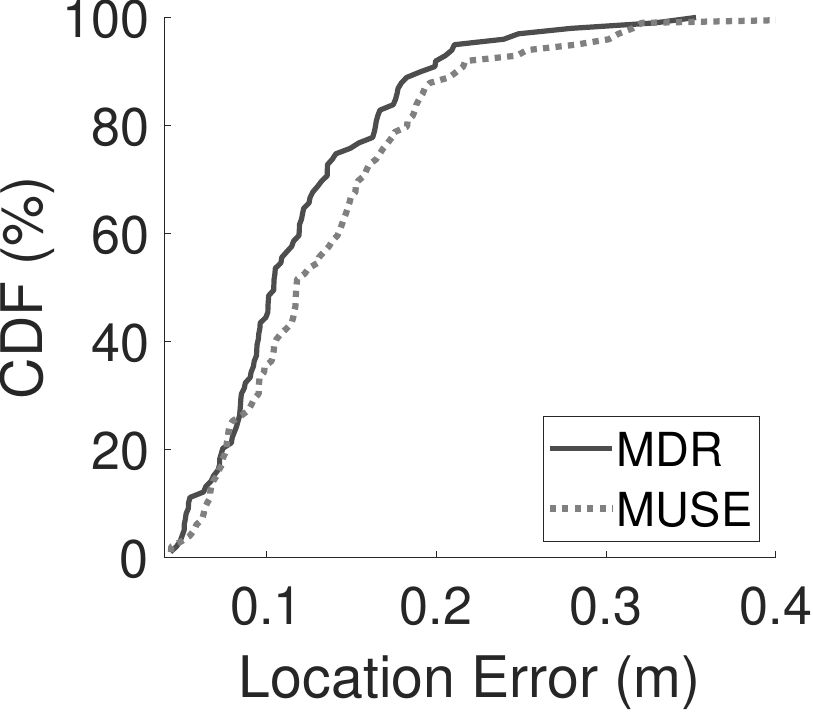}
        \vspace{-5mm}
        \caption{Overall location estimation error.}
        \label{localization}
    \end{minipage}
    \hfill
    \vspace{-2mm}
\end{figure*}

Additionally, \aliasSystem has an average orientation error of 17.73°, 5.39$\%$ lower than \textbf{\aliasSystem w/o AU} (18.74°).
This relative improvement is lower than the relative improvement in database accuracy (9.07$\%$), since magnetic calibration is only part of orientation estimation.
Reducing error in magnetic calibration only reduces part of the overall orientation error.
\highlight{\subsection{Database Auto-Updating Efficiency}}
\label{DBeval}
We now evaluate the efficiency of \aliasSystem building a database from scratch.
We first focus on the database size, namely the total amount of filled datapoint in the database, and the error of the database, namely the average direction error of all anchors in it.
As Figure \ref{build up} shows, in the first one minute, the database error rises swiftly, which is because the orientation error is usually zero after initialization and rises as the gyroscope drift accumulates.
In the next five minutes, the amount of datapoints gradually reaches the cap, and hardly increases.
This is probably because in the context of arm tracking, possible locations to be visited are limited.
As database keeps being updated, the database error gradually decreases, because the database voxels are visited by the smartwatch multiple times, as shown in Figure \ref{build up 2}.
In repeatedly visiting the locations, the updated anchors will aggregate into the datapoint at the same locations, resulting in a gradually decreasing database error, which eventually converges at 12.59°.
In conclusion, a database can adapt to new places with exploration of a few minutes, with adequate size and accuracy.

\highlight{
For new scenarios, \aliasSystem requires no pre-scanning or past database history.
Instead builds a new database from scratch in a few minutes.
If the magnetic environment changes, as long as it does not happen during the usage of \aliasSystem, \aliasSystem will be robust.
After the changes are complete, \aliasSystem will regard it as a new scenario and adapt to it within minutes.
\aliasSystem will fail if the environment changes constantly and significantly.
However, such scenarios are rare in real-life applications.
}

\subsection{Impact of Database Spatial Resolution} \label{secDatabaseResolution}
We examine the impact of the spatial resolution of the database, $l_{DB}$.
Figure \ref{prec} shows the orientation error of \aliasSystem with different $l_{DB}$, where $l_{DB}=0.1m$ returns the best performance.
If the resolution is too small, location error would overwhelm the resolution, resulting in anchors being updated in incorrect voxels.
Conversely, if the resolution is too large, the database would lose detailed information of magnetic distortion.


\begin{table}[ht]
    \centering
    \vspace{-1mm}
        \caption{Distortion-free scenario detection results.}
        \vspace{-2mm}
    \begin{tabular}{|l|c|c|c|}
        \hline
          & Criterion A & Criterion B & Using both \\
        \hline
         Precision & 92.31$\%$ & 84.75$\%$ & 96.00$\%$ \\
         \hline
         Recall & 96.00$\%$ & 100$\%$ & 96.00$\%$ \\
         \hline
         F1 score & 94.12$\%$ & 91.74$\%$ & 96.00$\%$ \\
         \hline
    \end{tabular}
  \vspace{-3mm}
    \label{use_or_not}
\end{table}


\subsection{Distortion-Free Detection Accuracy}
\label{detection_eval}
We identify scenarios with distortion magnitudes lower than 10° as distortion-free. 
According to Table \ref{places}, the database has limited contribution when distortion is lower than 10°. 
To avoid wasting computation, we designed two criteria to recognize these magnetic distortion-free scenarios and turn off the database in these cases. 
Here, we evaluate the effectiveness of these two criteria.

As Table \ref{use_or_not} shows, F1 score of criterion A is 94.12$\%$, and for B is 91.74$\%$.
If we combine A and B, we achieve an F1 score of 96$\%$, as well as 96$\%$ precision and 96$\%$ recall.
In conclusion, with the two criteria, we can conveniently and accurately identify distortion-free scenarios where database could be turned off to save computation.

In the 'Conference room', the database will decrease the accuracy by 1.83$\%$, due to potential database error.
If we turn off the database in these places, especially 'Conference room', we can save computation, while not affecting orientation accuracy much.
According to Table \ref{latency} in Section \ref{computation latency}, we can save 89.95$\%$ of the computation, as the database and its associated particle filter are not activated.



\subsection{Location Estimation} \label{eval_loc}

We also demonstrate the location estimation accuracy of \aliasSystem in Figure \ref{localization}.
\aliasSystem has an average error of 0.1201m, 12.61$\%$ lower than the existing state-of-the-art solution, MUSE.
This improvement mainly stems from the enhanced orientation accuracy, since particle filter uses it as input to track location.
Additionally, according to Figure \ref{spatial}, an error of 0.1201m does not significantly impact the magnetic field direction, thus ensuring accuracy in database query and updating.



\subsection{System Overhead}
\subsubsection{Computation Latency}\label{computation latency}
We implement \aliasSystem with the MATLAB Mobile App on the iPhone XS max, and evaluate its edge computing latency.
As the result, for every 1 second, \aliasSystem can process 3.37 seconds of IMU data to estimate orientation while simultaneously constructing a database and inferring location to support the database.
This demonstrates \aliasSystem's real-time computing capability on mobile devices.
For devices with lower computation resources, e.g., smartwatches, it is feasible to transmit the data to a smartphone using Bluetooth and perform computation on the smartphone.

We also examine the latency proportion of major system components of \aliasSystem, namely the complementary filter, particle filter and database.
As shown in Table \ref{latency}, particle filter consumes the majority of computation resources at 87$\%$, while complementary filter takes 10$\%$ and database takes less than 3$\%$.
Database operations are concise and efficient, resulting in its minimal proportion of computational consumption.

\begin{table}[ht]
    \centering
        \caption{Latency of major components in \aliasSystem.}
        \vspace{-2mm}
    \begin{tabular}{|l|c|c|c|}
        \hline
        Component & Complem. F. & Particle F. & Database \\
        \hline
        Relative latency & 10.05$\%$ & 87.08$\%$ &  2.87$\%$ \\
        \hline
    \end{tabular}
    \vspace{-2mm}
    \label{latency}
\end{table}

\subsubsection{Memory Overhead} \label{eval_memory}
We evaluate the memory occupation of the database.
Statistics on the iPhone XS max show that, for every 1m$^3$ of the real-world space, database takes an average of 21.63KB to store the magnetic anchors, with 0.1m-resolution. 
This demonstrates the scalability of \aliasSystem. 
The concise structure of the database, which is simply a 3-D array storing a 3-D vector in each of its units, contributes to this efficient memory usage.
\highlight{
\section{Real Application Case Study} \label{sec_case_study}
}

\highlight{ 
We deploy \aliasSystem into two real use cases to demonstrate its direct improvements for smart applications: star sky rendering on a smartphone, and in-air writing using a smartwatch.
Both applications rely solely on IMU sensors, and are conducted in magnetically distorted indoor environments.
}

\highlight{ \subsection{In-Air Writing} } \label{writing}

\begin{figure*} [ht]
    \vspace{-3mm}
    \subfigure['A']{\includegraphics[width=0.05\linewidth]{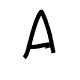}}
    \hfill
    \subfigure['B']{\includegraphics[width=0.05\linewidth]{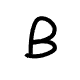}}
    \hfill
    \subfigure['C']{\includegraphics[width=0.05\linewidth]{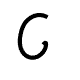}}
    \hfill
    \subfigure['D']{\includegraphics[width=0.05\linewidth]{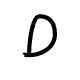}}
    \hfill
    \subfigure['E']{\includegraphics[width=0.05\linewidth]{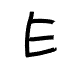}}
    \hfill
    \subfigure['F']{\includegraphics[width=0.05\linewidth]{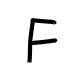}}
    \hfill
    \subfigure['G']{\includegraphics[width=0.05\linewidth]{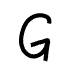}}
    \hfill
    \subfigure['H']{\includegraphics[width=0.05\linewidth]{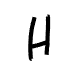}}
    \hfill
    \subfigure['I']{\includegraphics[width=0.05\linewidth]{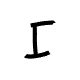}}
    \hfill
    \subfigure['J']{\includegraphics[width=0.05\linewidth]{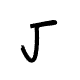}}
    \hfill
    \subfigure['K']{\includegraphics[width=0.05\linewidth]{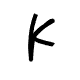}}
    \hfill
    \subfigure['L']{\includegraphics[width=0.05\linewidth]{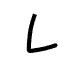}}
    \hfill
    \subfigure['M']{\includegraphics[width=0.054\linewidth]{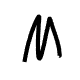}}
    \hfill
    \subfigure['N']{\includegraphics[width=0.05\linewidth]{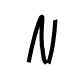}}
    \hfill
    \subfigure['1']{\includegraphics[width=0.05\linewidth]{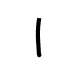}}
    \hfill
    \subfigure['2']{\includegraphics[width=0.05\linewidth]{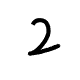}}
    \hfill
    \subfigure['3']{\includegraphics[width=0.05\linewidth]{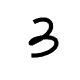}}
    \hfill
    \subfigure['4']{\includegraphics[width=0.05\linewidth]{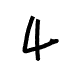}}
    \vspace{-4mm}
    \caption{In-Air Writing Trace Ground Truth Collected by VR System}
    \vspace{-1mm}
    \label{demo_gt}
    
    \subfigure['A']{\includegraphics[width=0.05\linewidth]{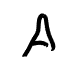}}
    \hfill
    \subfigure['B']{\includegraphics[width=0.05\linewidth]{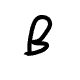}}
    \hfill
    \subfigure['C']{\includegraphics[width=0.05\linewidth]{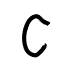}}
    \hfill
    \subfigure['D']{\includegraphics[width=0.05\linewidth]{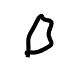}}
    \hfill
    \subfigure['E']{\includegraphics[width=0.05\linewidth]{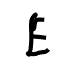}}
    \hfill
    \subfigure['F']{\includegraphics[width=0.05\linewidth]{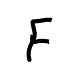}}
    \hfill
    \subfigure['G']{\includegraphics[width=0.05\linewidth]{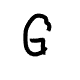}}
    \hfill
    \subfigure['H']{\includegraphics[width=0.05\linewidth]{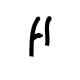}}
    \hfill
    \subfigure['I']{\includegraphics[width=0.05\linewidth]{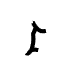}}
    \hfill
    \subfigure['J']{\includegraphics[width=0.05\linewidth]{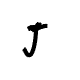}}
    \hfill
    \subfigure['K']{\includegraphics[width=0.05\linewidth]{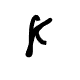}}
    \hfill
    \subfigure['L']{\includegraphics[width=0.05\linewidth]{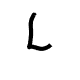}}
    \hfill
    \subfigure['M']{\includegraphics[width=0.054\linewidth]{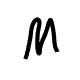}}
    \hfill
    \subfigure['N']{\includegraphics[width=0.05\linewidth]{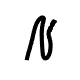}}
    \hfill
    \subfigure['1']{\includegraphics[width=0.05\linewidth]{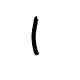}}
    \hfill
    \subfigure['2']{\includegraphics[width=0.05\linewidth]{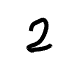}}
    \hfill
    \subfigure['3']{\includegraphics[width=0.05\linewidth]{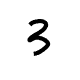}}
    \hfill
    \subfigure['4']{\includegraphics[width=0.05\linewidth]{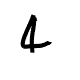}}
    \vspace{-4mm}
    \caption{In-Air Writing Trace Inferred by \aliasSystem}
    \vspace{-1mm}
    \label{demo_mdr}
    
    \vspace{1mm}
    \subfigure['A']{\includegraphics[width=0.05\linewidth]{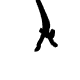}}
    \hfill
    \subfigure['B']{\includegraphics[width=0.05\linewidth]{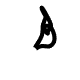}}
    \hfill
    \subfigure['C']{\includegraphics[width=0.05\linewidth]{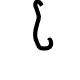}}
    \hfill
    \subfigure['D']{\includegraphics[width=0.05\linewidth]{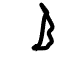}}
    \hfill
    \subfigure['E']{\includegraphics[width=0.05\linewidth]{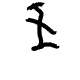}}
    \hfill
    \subfigure['F']{\includegraphics[width=0.05\linewidth]{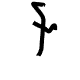}}
    \hfill
    \subfigure['G']{\includegraphics[width=0.05\linewidth]{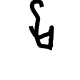}}
    \hfill
    \subfigure['H']{\includegraphics[width=0.05\linewidth]{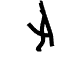}}
    \hfill
    \subfigure['I']{\includegraphics[width=0.05\linewidth]{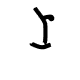}}
    \hfill
    \subfigure['J']{\includegraphics[width=0.05\linewidth]{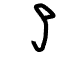}}
    \hfill
    \subfigure['K']{\includegraphics[width=0.05\linewidth]{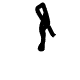}}
    \hfill
    \subfigure['L']{\includegraphics[width=0.05\linewidth]{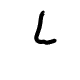}}
    \hfill
    \subfigure['M']{\includegraphics[width=0.054\linewidth]{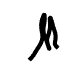}}
    \hfill
    \subfigure['N']{\includegraphics[width=0.05\linewidth]{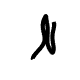}}
    \hfill
    \subfigure['1']{\includegraphics[width=0.05\linewidth]{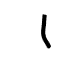}}
    \hfill
    \subfigure['2']{\includegraphics[width=0.05\linewidth]{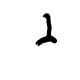}}
    \hfill
    \subfigure['3']{\includegraphics[width=0.05\linewidth]{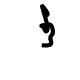}}
    \hfill
    \subfigure['4']{\includegraphics[width=0.05\linewidth]{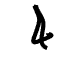}}
    \vspace{-4mm}
    \caption{In-Air Writing Trace Inferred by MUSE}
    \vspace{-1mm}
    \label{demo_muse}
\end{figure*}

\highlight{
We demonstrate an in-air writing application of \aliasSystem deployed on a smartwatch.
A potential real use case is that a user wears a smartwatch and draw symbols to interact with smart devices.
We write 20 symbols in the corridor (31° distortion), where magnetic distortion affects orientation estimation, and thus affects location estimation, causing inaccurate writing trace inference.
The symbols include letters, numbers, a triangle, and a rectangle, as shown in Figure \ref{demo_gt}, which are the ground truth writing traces collected by the VR device.
We test \aliasSystem and MUSE on the same arm motions, and show their inferred traces in Figure \ref{demo_mdr} and Figure \ref{demo_muse}.
We can see that \aliasSystem has relatively similar trace inference, while MUSE is not able to accurately track the writing.
This is because \aliasSystem resists magnetic distortion in the corridor and provides more accurate orientation results, which is necessary for tracking smartwatch location using IMU sensors.
}

\begin{figure*} [ht]
    \vspace{-5mm}
    \begin{minipage}[t]{0.21\linewidth}
        \hspace{-0.1cm}
        \centering
        \includegraphics[width=1\linewidth]{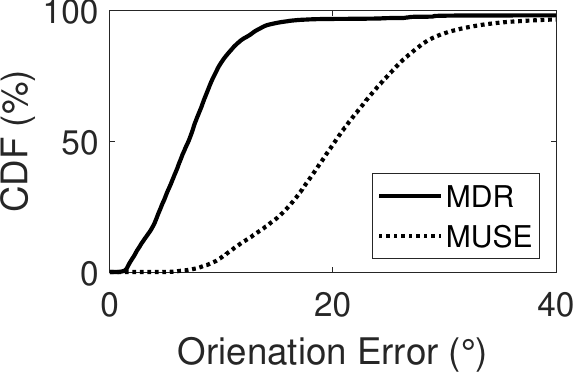}
        \vspace{-5mm}
       \caption{\small CDF of orientation estimation error during in-air writing.}
        \label{fig_demo_ori}
    \end{minipage}
    \hfill
    \begin{minipage}[t]{0.21\linewidth}
        \hspace{-0.1cm}
        \centering
        \includegraphics[width=1\linewidth]{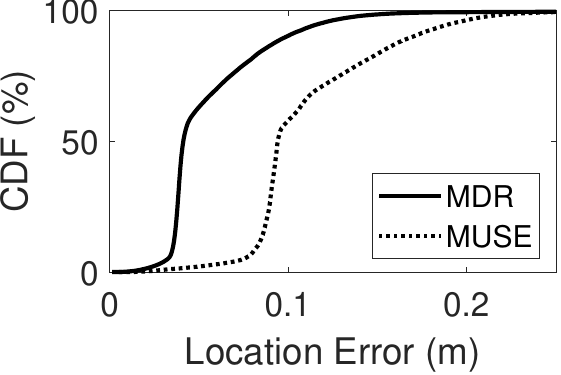}
        \vspace{-5mm}
       \caption{\small CDF of location estimation error during in-air writing.}
        \label{fig_demo_loc}
    \end{minipage}
    \hfill
    \begin{minipage}[t]{0.26\linewidth}
        \hspace{-0.1cm}
        \centering
        \includegraphics[width=1\linewidth]{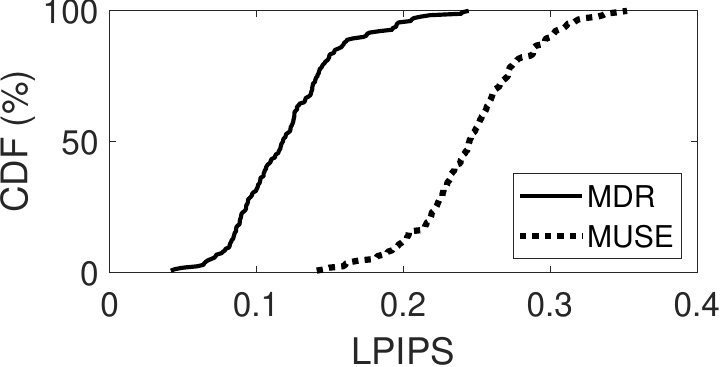}
        \vspace{-5mm}
       \caption{\small CDF of LPIPS of writing traces compared to ground truth traces.}
        \label{lpips}
    \end{minipage}
    \hfill
    \begin{minipage}[t]{0.25\linewidth}
        \hspace{-0.1cm}
        \centering
        \includegraphics[width=1\linewidth]{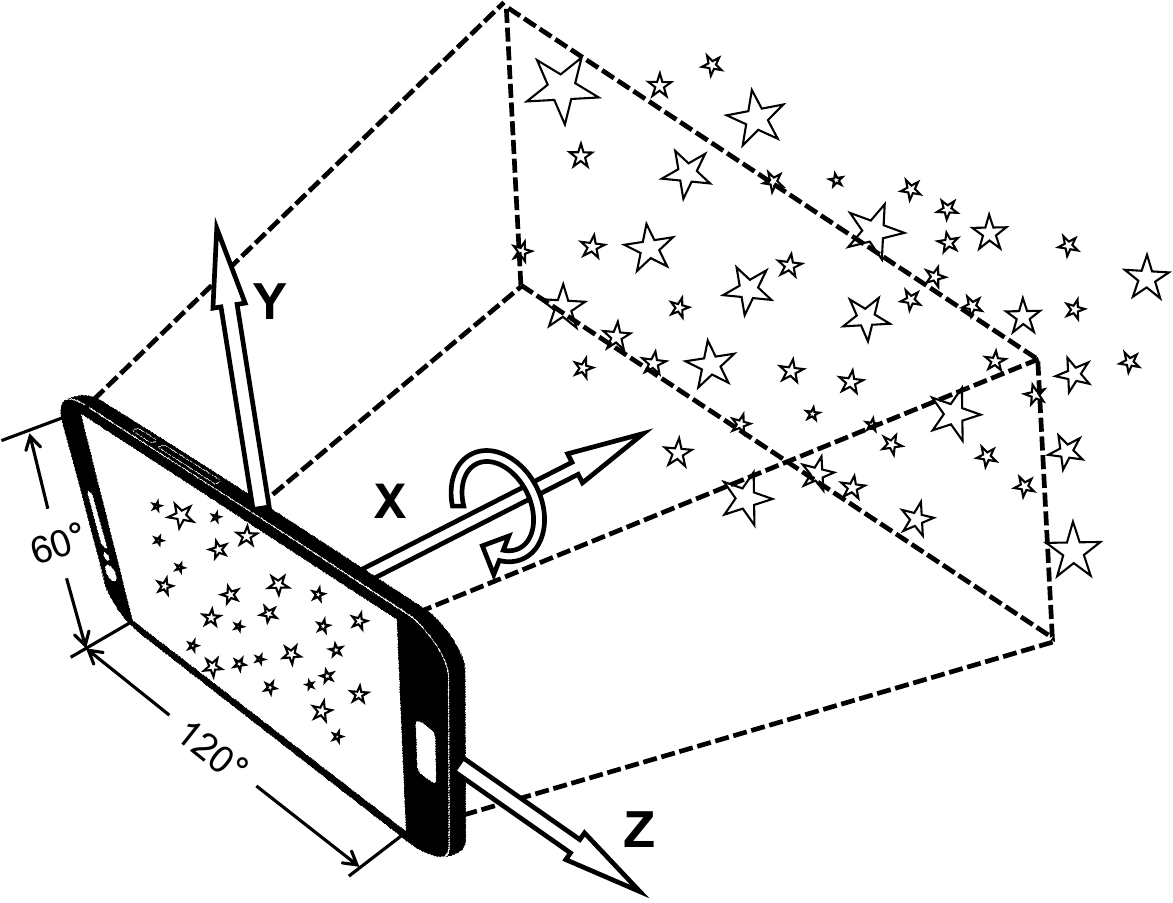}
        \vspace{-5mm}
       \caption{Star sky application illustration.}
        \label{ubi_star_sky}
    \end{minipage}
    \vspace{-2mm}
\end{figure*}

\highlight{
We write the symbols nine times, thus acquiring 180 samples.
The average orientation and location error during the writing are 8.27° and 0.056m for \aliasSystem, and for MUSE, 22.41° and 0.112m, with their CDFs shown in Figure \ref{fig_demo_ori} and Figure \ref{fig_demo_loc}.
We then use the popular LPIPS  (Learned Perceptual Image Patch Similarity) metric \cite{Zhang_2018_CVPR} to evaluate the similarity between the inferred writing traces and the ground truth traces.
The LPIPS metric is designed to be more aligned with human perception compared to traditional metrics like PSNR (Peak Signal-to-Noise Ratio) and SSIM (Structural Similarity Index).
Lower LPIPS values indicate better perceptual similarity between images. 
Figure \ref{lpips} shows the CDFs of the LPIPS of \aliasSystem and MUSE, both compared to the ground truth traces.
We can see that the LPIPS of \aliasSystem (with an average of 0.1216) is lower than MUSE (with an average of 0.2461), which shows that \aliasSystem better tracks the smartwatch, and provides more robust support for indoor smart applications.
}


\highlight{
\subsection{Star Sky Rendering} \label{eval_star}
}
Many star sky view applications are available on the market, e.g., Skyview, Star Walk, and Stellarium Mobile.
To demonstrate the direct benefit of orientation accuracy improvements, we deploy \aliasSystem into a star sky rendering application.
As an example scenario, a user uses a smartphone to look at constellations or stars.
The user does not really see the stars, but instead, the smartphone renders a simulated star sky based on the its orientation and the star map data.
However, metal furniture distorts the indoor magnetic field, making it hard for traditional methods to estimate accurate orientation and support accurate star sky rendering.
\highlight{
Inaccurate star sky rendering leads to impaired App user experience, and may also cause misleading guidance for astronomy fans.
For example, astronomy fans may use star sky Apps to find the approximate area of a constellation 
before precisely aligning the telescope, and thus inaccurate star sky information would decrease the efficiency of this coarse searching.
}



We use the Vizier star map data~\cite{vizier}, and randomly sample 10000 stars from it.
For every star, we have its direction relative to the Earth coordinate.
We use the smartwatch's orientation to simulate the view of a smartphone, as an intuitive example shown in Figure \ref{ubi_star_sky}.
Specifically, we use the X-axis of the WRF as the camera axis, and Y-axis as the vertical axis (up) of the view port, Z-axis (right) as horizontal axis of the view port.
For the view port, we set its horizontal view range as 120° and vertical range as 60°.
Note that, the direction of X-axis alone is not enough for rendering, as the rotation along X-axis also affects the of the view port. 
Instead, a 3-D orientation result is required to fully render the star sky.
We then project both the estimated orientation and the ground truth orientation onto the star map, and check which stars are included within the view ports.

\begin{figure}  [h] 
    \vspace{-1mm}
    \begin{minipage}[t]{0.47\linewidth}
        \centering
        \includegraphics[width=1\linewidth]{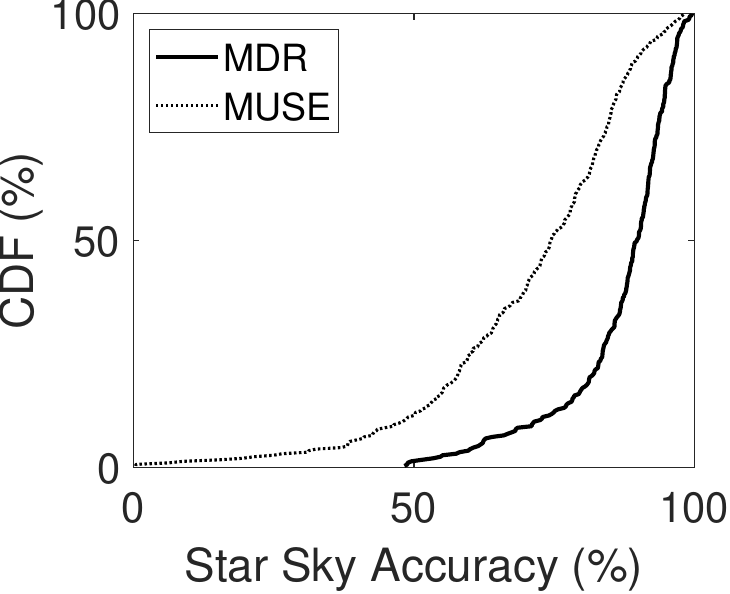}
        \vspace{-5mm}
       \caption{Star sky rendering accuracy CDF.}
        \label{star_2}
    \end{minipage}
    \hfill
    \begin{minipage}[t]{0.47\linewidth}
        \centering
        \includegraphics[width=1\linewidth]{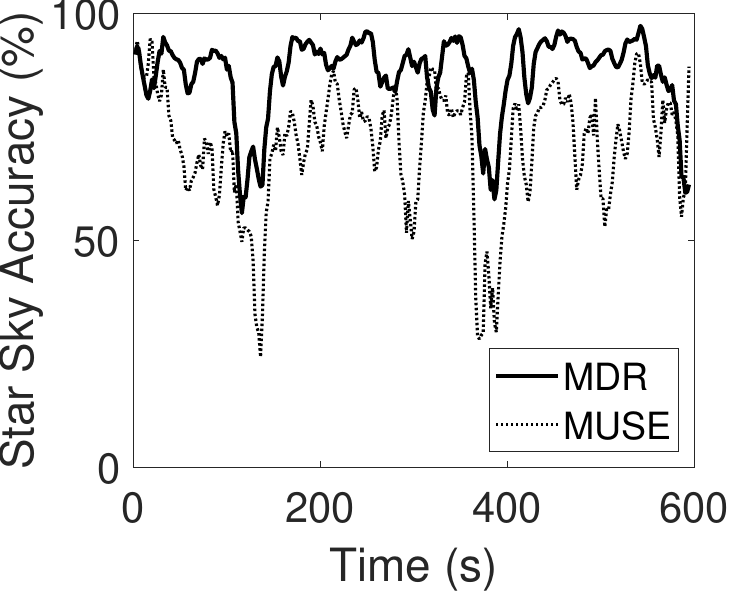}
        \vspace{-5mm}
        \caption{Star sky rendering accuracy over time.}
        \label{star_1}
    \end{minipage}
    \vspace{-3mm}
\end{figure}

We evaluate the rendering accuracy as 'how much percent of the stars in the ground truth view port are covered by the estimated view port'.
We use a data trace of 10 minutes collected in the Corridor (31° distortion) as example.
On this data trace, \aliasSystem has an average orientation error of 16.07°, and MUSE has an average error of 33.27°.

We deploy \aliasSystem and evaluate its star sky rendering accuracy, with MUSE as comparison.
The overall accuracy is 86.69$\%$ for \aliasSystem, and 70.92$\%$ for MUSE.
As shown in Figure~\ref{star_2}, we can see that \aliasSystem's accuracy hardly drops below 50$\%$.
As shown in Figure~\ref{star_1}, we can also see that \aliasSystem's accuracy is almost constantly better than MUSE.
This indicates that \aliasSystem can better support indoor application.


 

\section{Conclusion}\label{Conclusion}

We present \aliasSystem, a system that fundamentally models and resists magnetic distortion for orientation estimation. 
\aliasSystem constructs a database to model the magnetic distortion field and uses it to correct magnetic calibration.
The system automatically builds and updates the database, allowing it to adapt to new scenarios. 
Practical designs were made to improve database construction accuracy.
Extensive experiments demonstrate a significant improvement of \aliasSystem over existing methods in distorted scenarios, showcasing its effective distortion resistance.
\highlight{
\section{Appendix: Simulated Magnetic Distortion} \label{appendix}
}

\highlight{
To simulate different distortion levels, we generate data traces based on real arm motion data of one hour, and only modify the magnetometer to simulate different distortion levels.
To cancel the influence of other factors like motion pattern and motion speed, we generate data traces based on a large amount of data traces we collected in Section \ref{data}.
In this dataset, we know the ground truth of the smartwatch orientation and location by a ground truth data collection system developed in \cite{miao}.
We first transform the average magnetic field direction (usually north) from GRF into WRF, using the reverse of the ground truth orientation $\Theta_0^{-1}$ (acquired in Section \ref{ground truth}): 
$\overrightarrow{N}_{WRF}=\overrightarrow{N}_{GRF}\cdot\Theta_0^{-1}$.
Based on this magnetic field direction in WRF, and the original magnetometer measurement direction $\overrightarrow{M}$ (measurement in WRF), we calculate a linear combination: $k_c\cdot\overrightarrow{M} + (1-k_c) \overrightarrow{N}_{WRF}$, with a ratio $k_c$.
We normalize and use this new direction as the simulated magnetometer input.
Via tuning $k_c$, we can generate data with different distortion magnitude.
}


\end{document}